\documentclass[12pt,epsf,qsymbols]{article}
\pdfoutput=1
\usepackage[utf8]{inputenc}
\usepackage[normalem]{ulem}
\usepackage[english]{babel}
\usepackage{tabularx}
\usepackage{array}
\usepackage{graphics}
\usepackage[pdftex]{graphicx}
\usepackage{psfrag}
\usepackage{epsfig}
\usepackage{subfig}
\usepackage{amsmath}
\usepackage{mathtools}
\usepackage{amssymb}
\usepackage{bbold}
\usepackage{ulem}
\usepackage{setspace}
\usepackage{rotating}
\usepackage{colortbl}
\usepackage{longtable}
\usepackage{slashed}
\usepackage{braket}
\usepackage{lineno}
\makeatletter
\usepackage{textcomp}
\usepackage[usenames,dvipsnames]{xcolor}
\usepackage{relsize}
\usepackage{cite}

\usepackage{float}

\usepackage{bbm}
\usepackage{bm}
\usepackage{enumerate}
\usepackage{amsmath}
\usepackage{verbatim}

\usepackage{multirow}
\usepackage{arydshln}

\usepackage{hyperref}
\hypersetup{colorlinks,citecolor=nicegreen,linkcolor=niceblue}
\hypersetup{colorlinks=true}

\usepackage{pifont}

\allowdisplaybreaks

\setlongtables

\newcommand{\bea}{\begin{eqnarray}}
\newcommand{\eea}{\end{eqnarray}}
\newcommand{\beq}{\begin{equation}}
\newcommand{\eeq}{\end{equation}}
\newcommand{\ec}{\end{center}}
\newcommand{\bc}{\begin{center}}

\newcommand{\tev}{{\rm TeV}}

\newcommand{\pdir}{p\kern -5.2pt\raise 0.2ex\hbox {/}}

\newcommand{\vdir}{v\kern -5.75pt\raise 0.15ex\hbox {/}}
\newcommand{\kdir}{k\kern -5.75pt\raise 0.15ex\hbox {/}}
\newcommand{\epsdir}{\epsilon\kern -5.0pt\raise 0.15ex\hbox {/}}
\newcommand{\bvdir}{\bar{v}\kern -5.75pt\raise 0.15ex\hbox {/}}
\newcommand{\Ddir}{D\kern -7.75pt\raise 0.20ex\hbox {/}}
\newcommand{\Adir}{A\kern -7.75pt\raise 0.20ex\hbox {/}}
\newcommand{\ldir}{l\kern -5.0pt\raise 0.2ex\hbox{/}}
\newcommand{\varepsdir}{\varepsilon\kern -5.5pt\raise 0.15ex\hbox{/}}

\newcommand{\nn}{\nonumber}

\makeatother

\definecolor{niceblue}{rgb}{0.15,0.15,0.6}
\definecolor{nicegreen}{rgb}{0.1,0.5,0.1}
\definecolor{Red}{rgb}{1.,0.,0.}

\definecolor{Green}{rgb}{0.2,.7,0.2}

\begin{document}
\unitlength = 1mm

\thispagestyle{empty} 
\begin{flushright}
\begin{tabular}{l}
{\tt \footnotesize LPT-Orsay-19-03}\\
\end{tabular}
\end{flushright}
\begin{center}
\vskip 3.4cm\par
{\par\centering \textbf{\LARGE  
\Large \bf Probing low energy scalar leptoquarks  \\[0.3em] by the leptonic $W$ and $Z$ couplings }}
\vskip 1.2cm\par
{\scalebox{.85}{\par\centering \large  
\sc P.~Arnan$^a$, D.~Be\v{c}irevi\'c$^b$, F.~Mescia$^a$ and O.~Sumensari$^{c,d}$}
{\par\centering \vskip 0.7 cm\par}
{\sl 
$^a$~{Departament de F\'isica Qu\`antica i Astrof\'isica (FQA),
Institut de Ci\`encies del Cosmos (ICCUB), Universitat de Barcelona (UB), Spain.}}\\
{\par\centering \vskip 0.25 cm\par}
{\sl 
$^b$~Laboratoire de Physique Th\'eorique (B\^at.~210)\\
CNRS, Univ. Paris-Sud, Universit\'e Paris-Saclay, 91405 Orsay, France.}\\
{\par\centering \vskip 0.25 cm\par}
{\sl 
$^c$~Istituto Nazionale Fisica Nucleare, Sezione di Padova, I-35131 Padova, Italy}\\
{\par\centering \vskip 0.25 cm\par}
{\sl 
$^d$~Dipartamento di Fisica e Astronomia ``G.~Galilei", Università di Padova, Italy}\\

{\vskip 1.65cm\par}}
\end{center}

\vskip 0.85cm
\begin{abstract}
We compute the generic one-loop contribution involving scalar leptoquarks (LQ) to the $W$ and $Z$ leptonic decay widths. In our computation we include for the first time the finite terms and the corrections due to the external momenta of the electroweak bosons, which is a step beyond the leading-logarithmic approximation considered in the literature so far. We show that the terms we   include can be numerically quite significant. They amount to about $20\%$ for scalar LQ masses below $1.5$~TeV, as currently allowed by the direct searches at the LHC. To further illustrate the relevance of our results we revisit a model with two light scalar LQs, proposed to accommodate the $B$-physics anomalies. We show that the finite terms we computed can reduce the tension with the $Z$-pole data.
\end{abstract}
\newpage
\setcounter{page}{1}
\setcounter{footnote}{0}
\setcounter{equation}{0}
\noindent

\renewcommand{\thefootnote}{\arabic{footnote}}

\setcounter{footnote}{0}

\tableofcontents

\newpage

\section{Introduction}
\label{Sec:0}
Ever since the experimental observation that the lepton flavor universality (LFU) might be broken in the decays of $B$-mesons, there has been a great effort in the high energy physics community to build a model which would satisfactorily accommodate the breaking of LFU in such a way that a plethora of low energy physics observables remain consistent with the Standard Model (SM) and experiment. In addition to the low energy physics observables, any candidate model (theory) testable at the $\mathcal{O}(\tev)$ scale should also be consistent with the bounds deduced from the direct searches at the LHC. 
We remind the reader that the first departures from LFU have been observed in the decays which are the tree level $b\to c\ell\bar \nu$ processes in the SM, namely,~\footnote{Note that throughout this paper any lepton flavor will be denoted by $\ell \in \{e,\mu,\tau\}$, which is to be distinguished from the light leptons only, denoted by $l\in \{e,\mu\}$.}
\bea
\label{eq:RD_definition}
R_{D^{(\ast)}} = \left. \dfrac{\mathcal{B}(B\to D^{(\ast)} \tau\bar{\nu})}{\mathcal{B}(B\to D^{(\ast)} l \bar{\nu})}\right|_{l\in \{e,\mu\}},\qquad R_{J/\psi} =   \dfrac{\mathcal{B}(B_c\to J/\psi \tau\bar{\nu})}{\mathcal{B}(B_c\to J/\psi \mu \bar{\nu})}.
\eea
The main benefit of considering this kind of ratios is that the Cabibbo--Kobayashi-Maskawa (CKM) couplings and a considerable amount of hadronic uncertainties cancel so that the deviations of the ratios from their SM estimate can be clearly interpreted as signs of the LFU violation (LFUV). More specifically, the most recent average HFLAV values~\cite{Amhis:2016xyh} read: 
\begin{align}
R_{D}^\mathrm{exp} &=0.41(5), &  R_{D}^\mathrm{SM} &=0.300(8), \nn\\
R_{D^\ast}^\mathrm{exp} &=0.31(2),  &R_{D^\ast}^\mathrm{SM} &=0.257(3), \\
R_{J/\psi}^\mathrm{exp} &=0.71(25), & R_{J/\psi}^\mathrm{SM} &\simeq 0.23(1).\nn
\end{align}
Even though the $5\sigma$ departure from LFU is still missing, this astonishing results require a plausible explanation which should also verify a conservative bound $\mathcal{B}(B_c\to \tau \bar{\nu}) \lesssim 30\%$~\cite{Li:2016vvp,Alonso:2016oyd}.
The puzzle regarding the LFUV in $B$-decays become even more intriguing after LHCb considered the $b\to s\l^+l^-$ decays, which are loop-induced in the SM, and measured 
\bea
R_{K^{(\ast)}}^{[q_1^2, q_2^2]} =  \dfrac{\mathcal{B}'(B\to K^{(\ast)} \mu\mu)}{\mathcal{B}'(B\to K^{(\ast)} ee)}  \,,
\label{eq:RK_definition}
\eea
where $\mathcal{B}'$ stands for the partial branching fraction, integrated between $q_1^2$ and $q_2^2$.
A comparison between the measured values~\cite{Amhis:2016xyh}, and those computed in the SM~\cite{Bordone:2016gaq}:
\begin{align}
R_K^\mathrm{exp}\equiv &R_{K^{+}}^{[1,6]}=0.75(9),  & & R_{K^{+}}^{[1,6]\,\mathrm{SM}} = 1.00(1), \nn \\ 
R_{K^\ast}^\mathrm{exp}\equiv &R_{K^{\ast 0}}^{[1.1,6]} =0.71(10), & &  R_{K^{\ast 0}}^{[1.1,6]\,\mathrm{SM}}  = 1.00(1),  \\ 
&R_{K^{\ast 0}}^{[0.045,1.1]} =0.68(10),  & & R_{K^{\ast 0}}^{[0.045,1.1]\,\mathrm{SM}}  =0.98(1),\nn
\end{align}
shows that the measured values are systematically about $2.5 \sigma$ lower than predicted.

Building a model which can describe both of these anomalies is difficult because the two kinds of processes are probing different new physics scales~\cite{DiLuzio:2017chi}. 
Furthermore, requiring the compatibility with a number of measured observables at low energies allows one to eliminate many candidate models. For that reason most of the theoretical work has been focused in reconciling theory with experiment for either $R_{D^{(\ast)}} $ or $R_{K^{(\ast)}}$, and only a few attempts have been successful in accommodating both types of the above-mentioned ``{\it $B$-physics anomalies}". 
Among the specific models proposed so far those involving the leptoquark states (LQ's) seem to be the most appealing. They can modify the SM prediction through a tree-level or one-loop diagrams, depending on the structure of Yukawa couplings chosen to make the model consistent with the low- and high-energy observables. The most recent assessment of viability of the single LQ models has been made in Ref.~\cite{Angelescu:2018tyl}. It appears that none of the scalar LQ mediators alone can be used to accommodate both kinds of anomalies. In such a situation one should either combine two different scalar LQ's or opt for the vector LQ's. The latter option is difficult to implement in a minimalistic scenario because the theory in which the SM is extended by only the vector LQ's at $\mathcal{O}(1-10\,\tev)$ is not renormalizable. To render the loop corrections finite in such a scenario one would need to specify the ultra-violet completion of the theory,  which in turn involves a number of new parameters (cf. Refs.~\cite{DiLuzio:2017vat,Crivellin:2018yvo,Blanke:2018sro,DiLuzio:2018zxy,Bordone:2017bld,Bordone:2018nbg}). For that reason and throughout this work we will focus on the scalar LQ's only. The attempts to combine two LQ's and accommodate the $B$-physics anomalies have been proposed in Refs.~\cite{Buttazzo:2017ixm,Marzocca:2018wcf,Crivellin:2017zlb,Becirevic:2018afm}.

We emphasize once again that a viability of any proposed scenario can be assessed through a careful comparison between theory and experiment for a number of low- and high-energy observables. For most quantities in the scenarios of new physics in which the SM is extended by one or more LQ's, the expressions can be found in Ref.~\cite{Dorsner:2016wpm}. It appears, however, that the expressions for the leptonic decays of $Z$ and $W$ bosons are not available. These processes, together with the leptonic $\tau$-decays, are the subject of this paper. The leading LQ contributions modify the decay rates of all these processes at the one-loop level. We compute them for all of the possible scalar LQ scenarios with the most general structure of Yukawa couplings.~\footnote{The term ``{\it Yukawa couplings}", in this paper, is used to designate the couplings among a quark, a lepton and a scalar LQ.} The necessity for checking whether or not a proposed model is consistent with the electro-weak precision tests is, of course, not new. Its importance has been stressed in Ref.~\cite{Feruglio:2016gvd} where the leading logarithmic contribution has been used  for the renormalization group running from $m_{W,Z}$ to $\mathcal{O}(1\,\tev \div 10\,\tev)$, scales at which a given LQ is supposed to be on its mass shell. Since the $\tev$ scale and $m_{W,Z}$ are not far too apart from each other, checking on the finite LQ contributions to $Z$ and $W$ decays becomes important. The corresponding results are provided in this paper. 

We will illustrate the importance of inclusion of the finite contributions to $\mathcal{B}(Z\to \ell\ell)$ and $\mathcal{B}(W\to \ell\bar \nu)$ arising in some specific LQ models proposed in the literature so far. Moreover, we will reexamine a concrete model proposed to accommodate the $B$-physics anomalies by considering a singlet ($S_1$) and triplet ($S_3$) LQ states with couplings to left-handed SM fermions. We will show that the corrections we compute here reduce the tension observed for this model between the $Z$-pole data and the deviations in $R_{D^{(\ast)}}^{\mathrm{exp}}$.  

\

The remainder of this paper is organized as follows: In Sec.~\ref{Sec:1} we recall the scalar LQ representations and their most general Yukawa couplings allowed by the SM gauge symmetry. In Sec.~\ref{Sec:2} and Sec.~\ref{Sec:3} we describe our general computation of LQ contributions to $Z$ and $W$ decays into leptons. Finally, in Sec.~\ref{Sec:4} we illustrate the relevance of our results to the concrete phenomenological situation on a model proposed to explain the $B$-physics anomalies. We shortly conclude in Sec.~\ref{Sec:C}.

\section{Scalar Leptoquarks\label{Sec:1}}

In this Section we remind the reader of the scalar LQ Lagrangians. We follow the notation of Ref.~\cite{Dorsner:2016wpm} and specify LQs by their SM quantum numbers, $(SU(3)_c,SU(2)_L)_Y$. In this way the electric charge, $Q=Y+I_3$, is the sum of the hypercharge ($Y$) and the third component of the weak isospin ($I_3$). In the left-handed doublets, $Q_i=[(V^\dagger u_L)_i~d_{L\,i}]^T$ and $L_i=[(U\nu_L)_i~\ell_{L\,i}]^T$, 
the matrices $V$ and $U$ are respectively the CKM and the Pontecorvo--Maki--Nakagawa--Sakata (PMNS) matrices. As the neutrino masses are insignificant for 
phenomenology of this paper we can set $U= \mathbb{1}$.  
\begin{align}
\begin{split}
\label{eq:yuk-R2}
R_2= \left( \mathbf{3}, \mathbf{2}\right)_{7/6}:\qquad \mathcal{L}_{R_2}=& -\left( y_{R_2}^L\right)_{ij}\, \bar{u}_{R_i} R_{2} i\tau_2 L_{j} + \left( y_{R_2}^R\right)_{ij} \bar{Q}_i R_{2} \ell_{R_j} +\textrm{h.c.}\,, \\
\end{split}\\[0.8em]
\begin{split}
\label{eq:yuk-R2tilde}
\widetilde R_2= \left( \mathbf{3}, \mathbf{2}\right)_{1/6}:\qquad \mathcal{L}_{\widetilde{R}_2}=&- \left(y_{\widetilde{R}_2}^L\right)_{ij}\, \bar{d}_{R_i} \widetilde{R}_2  i\tau_2 L_j+\left( y_{\widetilde{R}_2}^R\right)_{ij}\, \bar{Q}_i \widetilde{R}_2 \nu_{R_j} +\textrm{h.c.}\,, \\
\end{split}\\[0.8em]
\begin{split}
\label{eq:yuk-S1}
S_1= \left( \overline{\mathbf{3}}, \mathbf{1}\right)_{1/3}:\qquad  \mathcal{L}_{S_1}=& \left( y_{S_1}^L\right)_{ij}\, \bar{Q}^C_i i\tau_2 S_{1} L_{j}+  \left( y_{S_1}^R\right)_{ij}\, \bar{u}^C_{R_i}  S_{1} \ell_{R_j}\\[0.4em]&\hspace*{8em}+\left( y_{S_1}^{\prime\,R}\right)_{ij}\,\bar{d}_{Ri}^{\,C} S_1\nu_{Rj} + \textrm{h.c.}\,,\\
\end{split}\\[0.8em]
\begin{split}
\label{eq:yuk-S3}
S_3= \left( \overline{\mathbf{3}}, \mathbf{3}\right)_{1/3}:\qquad  \mathcal{L}_{S_3}=& \left( y_{S_3}^L\right)_{ij}\,  \bar{Q}^C_i i\tau_2 (\vec{\tau} \cdot \vec{S}_{3})  L_{j}+\textrm{h.c.}\,,\\
\end{split}\\[0.8em]
\begin{split}
\label{eq:yuk-S1t}
\widetilde S_1= \left( \overline{\mathbf{3}}, \mathbf{1}\right)_{4/3}:\qquad \mathcal{L}_{\widetilde{S}_1}=& \left( y_{\widetilde{S}_1}^R \right)_{ij}\,  \bar{d }^{\,C}_{R_i}\widetilde{S}_1 \ell_{R_j} +\textrm{h.c.}\,,\\
\end{split}\\[0.8em]
\begin{split}
\label{eq:yuk-S1bar}
\bar S_1= \left( \overline{\mathbf{3}}, \mathbf{1}\right)_{-2/3}:\qquad \mathcal{L}_{\overline{S}_1}=&  \left(  y_{\bar{S}_1}^R  \right)_{ij}\,  \bar{u}_{R_i}^{C}\, \bar{S}_1 \nu_{R_j} +\textrm{h.c.}\,,\\
\end{split}
\end{align}
where, as usual, the fermion fields $\psi_{L,R}=P_{L,R}\psi$ with  $P_{L,R}=(1 \mp \gamma^5)/2$, and $\psi^C$ stands for a charge conjugated fermion, while $\tau_k$ denote the Pauli matrices. 
Note that we neglected the LQ couplings to diquarks in Eqs.~(\ref{eq:yuk-S1}-\ref{eq:yuk-S1bar}) which is necessary for 
stability of the proton~\cite{Dorsner:2016wpm}. $y^{L,R}_\mathrm{LQ}$ are the matrices of Yukawa couplings the components of which correspond to the quark and lepton indices in the weak interaction eigenbasis. It is often useful to label the scalar leptoquarks through its flavour number $F=3B+L$, where $B$ and $L$ stand for the baryon and the lepton number respectively. In that way $R_2$ and $\widetilde{R}_2$ are $F=0$ leptoquarks while $S_1,S_3,\widetilde{S}_1$ and $\bar{S}_1$ are $|F|=2$ leptoquarks.

For phenomenological considerations it is more convenient to work in the mass eigenbasis. After absorbing the matrices of rotation to the mass eigenstate basis into the redefinition of Yukawa matrices ($y^{L,R}_\mathrm{LQ}$), and by accounting for the usual CKM mixing matrix $V$, we can write
\begin{align}
\begin{split}
\mathcal{L}_{R_2}=&-  \left( y_{R_2}^L\right)_{ij} \bar{u}_i P_L \ell_j  \, R_2^{(5/3)}+ \left(  y_{R_2}^L \right)_{ij} \bar{u}_i P_L \nu_j  \, R_2^{(2/3)}\\
&+ \left( V y_{R_2}^R\right)_{ij} \bar{u}_i P_R \ell_j  \, R_2^{(5/3)}+ \left( y_{R_2}^R \right)_{ij}\bar{d}_i P_R \ell_j  \, R_2^{(2/3)}  + \text{h.c.}\;,\nonumber\\
\end{split}\\[0.8em]
\begin{split}
\mathcal{L}_{\widetilde{R}_2}=&-  \left( y_{\widetilde{R}_2}^L \right)_{ij}  \bar{d}_i P_L \ell_j  \, \widetilde{R}_2^{(2/3)}+  \left( y_{\widetilde{R}_2}^L \right)_{ij}  \bar{d}_i P_L \nu_j \,  \widetilde{R}_2^{(-1/3)}\\
&+ \left( V y_{\widetilde{R}_2}^R\right)_{ij}  \bar{u}_i P_R \nu_j \,  \widetilde{R}_2^{(2/3)}+ \left( y_{\widetilde{R}_2}^R \right)_{ij}  \bar{d}_i P_R \nu_j  \, \widetilde{R}_2^{(-1/3)}  + \text{h.c.}\;,\nonumber\\
\end{split}\\[0.8em]
\begin{split}
\mathcal{L}_{S_1}=&-\left(y_{S_1}^L \right)_{ij}  \bar{d}^{\,C} P_L \nu  \, S_1+\left( V^* y_{S_1}^L \right)_{ij}  \bar{u}^C_i P_L \ell_j  \, S_1\\
&+ \left(y_{S_1}^R \right)_{ij}  \bar{u}^C_i P_R \ell_j  \, S_1+\left(y_{S_1}^{
\prime\,R} \right)_{ij} \bar{d}^{\,C}_i P_R \nu_j \,  S_1  + \text{h.c.}\;,\nonumber\\
\end{split}\\[0.8em]
\begin{split}
\mathcal{L}_{S_3}=&- \left(y_{S_3}^L  \right)_{ij}  \bar{d}^{\,C}_i P_L \nu_j  \, S_3^{(1/3)}-\sqrt{2} \left( y_{S_3}^L  \right)_{ij}  \bar{d}^{\,C}_i P_L \ell_j  \, S_3^{(4/3)}\nonumber\\
&+\sqrt{2} \left(V^*y_{S_3}^L  \right)_{ij}  \bar{u}^C_i P_L\nu_j  \, S_3^{(-2/3)}-\left(V^* y_{S_3}^L  \right)_{ij}  \bar{u}^C_i P_L \ell_j  \, S_3^{(1/3)}  + \text{h.c.}\;.\\
\end{split}\\[0.8em]
\begin{split}
\mathcal{L}_{\widetilde{S}_1}=& \left( y_{\widetilde{S}_1}^R  \right)_{ij} \bar{d}^{\,C}_i P_R \ell_j  \, \widetilde{S}_1+\textrm{h.c.}\;,\nn \\
\end{split}\\[0.8em]
\begin{split}
&\mathcal{L}_{\bar{S}_1}=   \left( y_{\bar{S}_1}^R  \right)_{ij} \bar{u}^{C}_i P_R \nu_j \,\bar{S}_1+ \text{h.c.}\;,\nn
\end{split}
\end{align}
where in the superscript of the non-singlet LQ field we note the component corresponding to the specific electric charge eigenstate which we assume to be mass degenerate.
We stress once again that we set the PMNS matrix to $U= \mathbb{1}$.

\section{Leptoquark contributions to $Z\to\ell\ell$}
\label{Sec:2}

\subsection{Effective field theory description}

\begin{figure}[t!]
\centering
\includegraphics[width=0.25\textwidth]{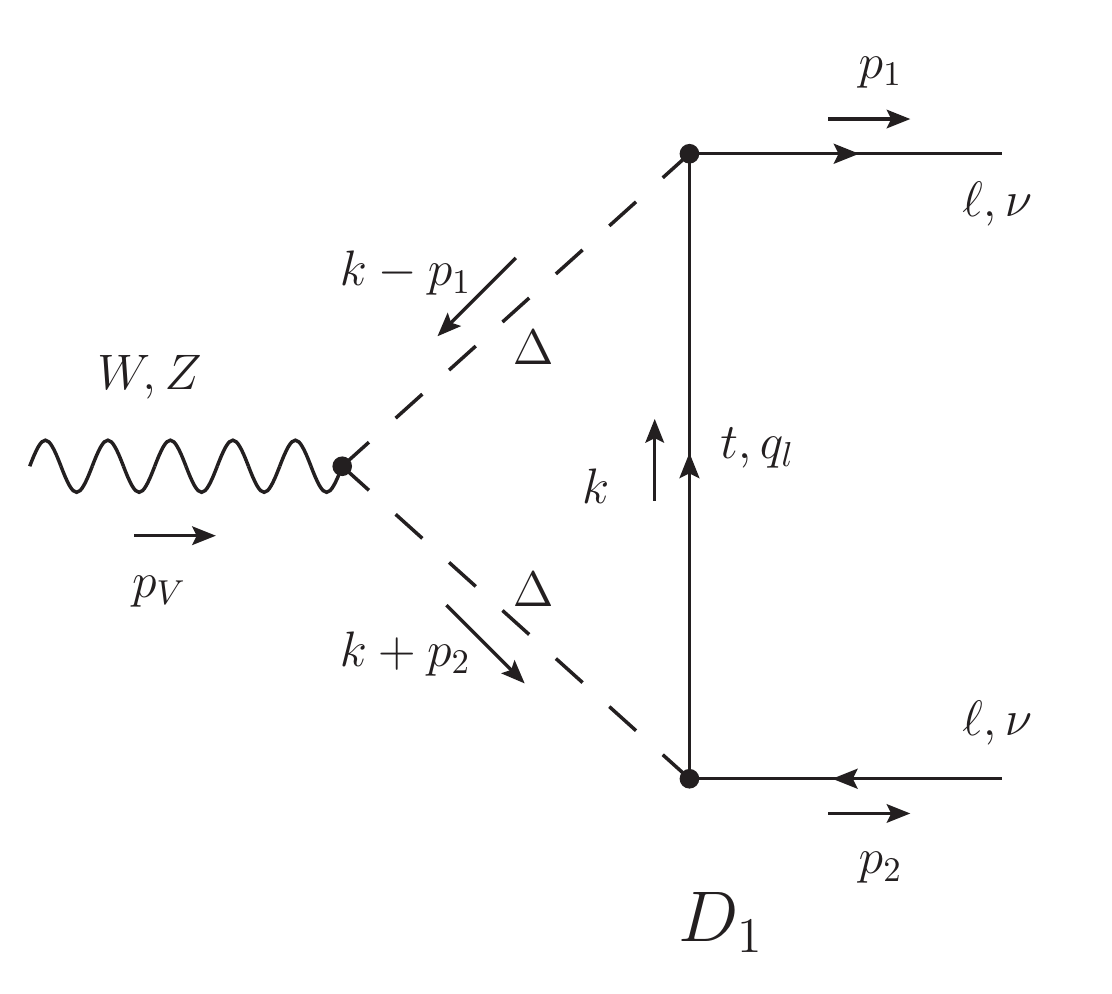}
\includegraphics[width=0.25\textwidth]{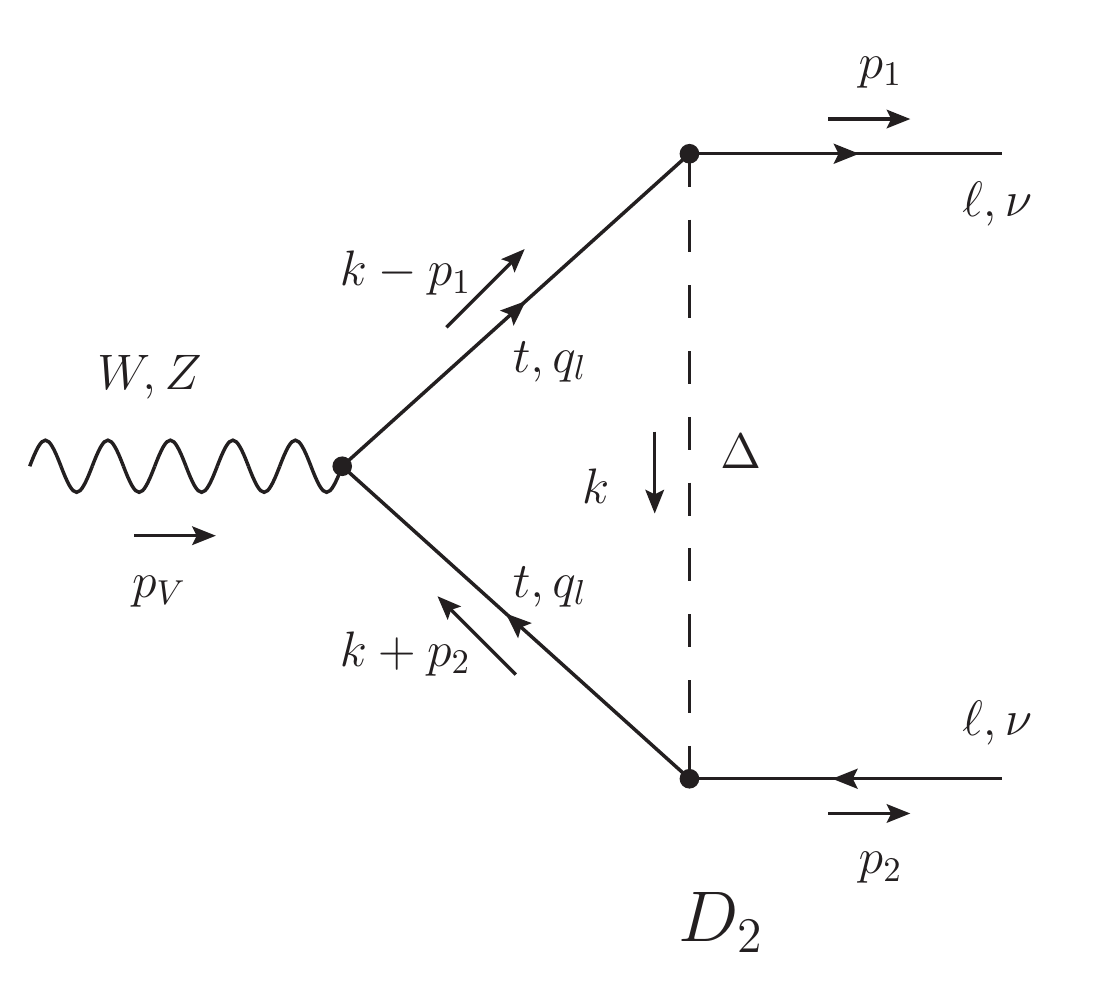}
\includegraphics[width=0.20\textwidth]{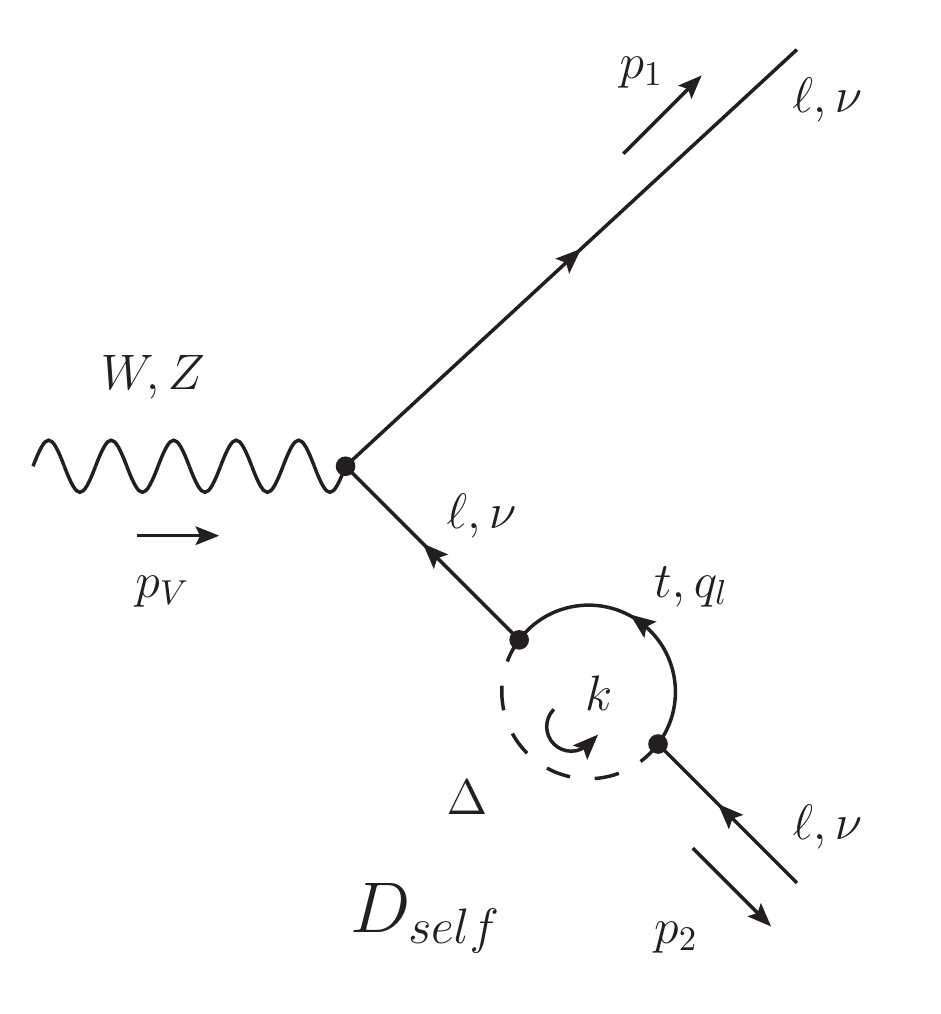}
\includegraphics[width=0.20\textwidth]{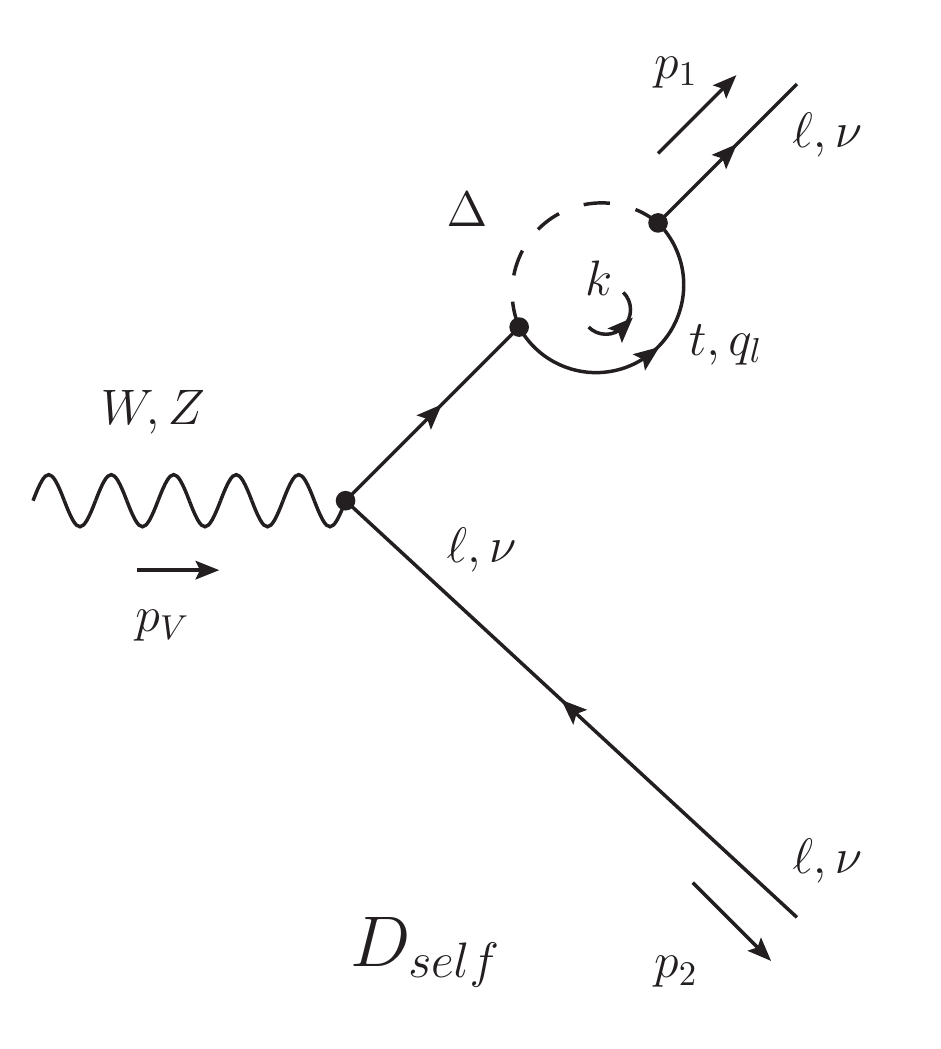}
\caption{\small\sl Scalar LQ ($\Delta$) contributions at one-loop to $Z\to \ell_i \ell_j$, $Z\to \nu_i \nu_j$ and $W\to \ell_j \nu_i$  .}
\label{fig:diagrams-Z}
\end{figure}

Leptoquarks contribute to the $Z$ couplings to leptons via the loop diagrams illustrated in Fig.~\ref{fig:diagrams-Z}. The effective Lagrangian describing the $Z$-boson interaction to generic fermions $f_{i,j}$ can be written as
\begin{align}
\label{eq:leff-z}
\delta \mathcal{L}_{\mathrm{eff}}^Z = \dfrac{g}{\cos \theta_W} \sum_{f,i,j} \bar{f}_{i} \gamma^\mu
\Big{[}g_{f_L}^{ij} \, P_L+ g_{f_R}^{ij} \,P_R \Big{]} f_j\, Z_\mu\end{align}
\noindent where $g$ is the $SU(2)_L$ gauge coupling, $\theta_W$ is the Weinberg angle, and
\begin{align}
g_{f_{L(R)}}^{ij} &= \delta_{ij}\,g_{f_{L(R)}}^{\mathrm{SM}}+\delta g_{f_{L(R)}}^{ij}\,,
\end{align}
 
\noindent with $g_{f_L}^{\mathrm{SM}}=I_3^f-Q^f\sin^2\theta_W$ and $g_{f_R}^{\mathrm{SM}}=-Q^f\sin^2\theta_W$. LQ loop contributions are described by the effective coefficients $\delta g_{f_{L(R)}}^{ij}$. The corresponding $Z$-boson branching fractions are then given by \footnote{For $i\neq j$ the computation of the branching ratio has to be interpreted as the average $\frac{1}{2}\left[\mathcal{B}(Z\to f_i \bar{f_j})+\mathcal{B}(Z\to f_j \bar{f_i})\right]$}
\begin{align}
\begin{split}
\mathcal{B}(Z\to f_i \bar{f_j}) = \dfrac{m_Z \lambda^{1/2}_Z}{6\pi v^2 \Gamma_Z} \Bigg{[}&\left(|g_{f_L}^{ij}|^2+|g_{f_R}^{ij}|^2\right)\Bigg{(}1- \dfrac{m_{i}^2+m_{j}^2}{2m_Z^2}- \dfrac{(m_i^2-m_j^2)^2}{2m_Z^4}\Bigg{)} \\
&+ 6 \dfrac{m_i m_j}{m_Z^2}\, \mathrm{Re}\left[g_{f_L}^{ij} \left( g_{f_R}^{ij}\right)^\ast\right] \Bigg{]}\,,
\end{split}
\end{align}
where $m_{{i,j}}$ are the fermion masses and $\lambda_Z\equiv[m_Z^2-(m_i-m_j)^2][m_Z^2-(m_i+m_j)^2]$.~\footnote{This expression also applies to the decays $Z\to \nu\nu$ if neutrinos are assumed to be Dirac particles. If lepton number is violated, this expression should be modified. Both formulas, however, agree in the limit $m_{i,j}\to 0$. See Ref.~\cite{Abada:2016plb} for a similar discussion in the case of $K\to \pi \nu \bar{\nu}$ decays.}  Contributions to these rates are constrained by the measurement of both flavor conserving and flavor violating $Z$ decays at LEP~\cite{Tanabashi:2018oca}. In particular, LEP measured the effective couplings~\cite{ALEPH:2005ab}

\begin{equation}
\begin{alignedat}{2}
g_V^{e,\,\mathrm{exp}} &= -0.03817(47)\,, \qquad\qquad\quad &g_A^{e,\,\mathrm{exp}} =&-0.50111(35)\,,\\[0.3em]
g_V^{\mu,\,\mathrm{exp}} &= -0.0367(23)\,, &g_A^{\mu,\,\mathrm{exp}}=&-0.50120(54)\,,\\[0.3em]
g_V^{\tau,\,\mathrm{exp}} &= -0.0366(10)\,, &g_A^{e,\,\mathrm{exp}}=&-0.50204(64)\,,
\end{alignedat}
\end{equation}

\noindent which are related to the couplings in Eq.~\eqref{eq:leff-z} via the relations $g_{V(A)}^{ij}=g_{\ell_L}^{ij}\pm g_{\ell_R}^{ij}$. Note that for $i=j$ we simplify the notation by dropping one superindex. Another important observable is the effective number of neutrinos~\cite{ALEPH:2005ab}

\begin{equation}
N_\nu^{\mathrm{exp}} = 2.9840(82)\,,
\end{equation}

\noindent which will constraint the LQ couplings to neutrinos via~\cite{ALEPH:2005ab}

\begin{equation}
N_\nu = \sum_{i,j} \left[ \left|\delta_{ij}+\dfrac{\delta g_{\nu_L}^{ij}}{g_{\nu_L}^{\mathrm{SM}}}\right|^2+\left|\dfrac{\delta g_{\nu_R}^{ij}}{g_{\nu_L}^{\mathrm{SM}}}\right|^2\right]\,,
\end{equation}

\noindent where $i,j\in \lbrace e,\mu,\tau \rbrace$ and neutrino masses have been neglected.

\subsection{One-loop matching}

We shall now provide an expressions for the couplings $\delta g^{ij}_{L(R)}$ for each of the leptoquark models listed in Sec.~\ref{Sec:1}. We focus our discussion onto the leptonic $Z$ couplings since these are the most precisely determined by experiment. Our discussion can be adapted ``\textit{mutatis mutandis}" to the $Z$ couplings to quarks.  Before presenting our results, we define our convention for the covariant derivative as

\begin{equation}
D_\mu = \partial_\mu + i g^\prime\, Y B_\mu-i g\, I^k W_\mu^k-i g_S\, T^A G_\mu^A\,,
\end{equation}

\noindent where $Y$ is the hypercharge, and $T^A$ and $I^k$ are the relevant $SU(3)_c$ and $SU(2)_L$ generators, respectively. After the electroweak symmetry breaking, this expression can be rewritten as

\begin{equation}
\label{eq:cov-dev}
D_\mu = \partial_\mu - i \dfrac{g}{\sqrt{2}}\, (I^+ W_\mu^+ +I^- W_\mu^-)-i \dfrac{g}{\cos \theta_W}\,  (I_3-Q \sin^2 \theta_W)Z_\mu+i e\, Q A^\mu-i g_S\, T^A G_\mu^A\,,
	\end{equation}

\noindent where $e=g \sin \theta_W=g^\prime \cos \theta_W$, $Q=Y+I_3$ and $I^\pm=(I_1\pm i\, I_2)$, as usual. To present our results in a compact form, we consider a general Yukawa Lagrangian defined by

\begin{align}
\label{eq:yuk-F0}
\mathcal{L}^{F=0}_{\mathrm{yuk.}} &= \bar{q}_i\big{[}l_{ij} P_R + r_{ij}P_L\big{]} \ell_j \, \Delta + \mathrm{h.c.}\\[0.4em]
\label{eq:yuk-F2}
\mathcal{L}^{F=2}_{\mathrm{yuk.}} &= \bar{q}_i^{\,C}\big{[}l_{ij} P_R + r_{ij}P_L\big{]} \ell_j \, \Delta + \mathrm{h.c.}
\end{align}

\noindent where $q$ and $\ell$ are generic quark and lepton flavors,  $l_{ij}$ and $r_{ij}$ denote the generic Yukawa couplings, and $\Delta$ is a leptoquark mass eigenstate, which belongs to one of the $SU(2)_L$ multiplets listed in Eq.~\eqref{eq:yuk-R2}--\eqref{eq:yuk-S1bar}. Our results will be presented in such a way that the expression for a specific model can be obtained by simply comparing Eqs.~(\ref{eq:yuk-F0},\ref{eq:yuk-F0}), to the Yukawa lagrangians listed in Sec.~\ref{Sec:1}. In this way, one can determine $l_{ij}$ and $r_{ij}$ for each leptoquark charge eigenstate contributing to $Z\to \ell\ell$ or $Z\to \nu \bar{\nu}$, which should then be summed up to give the final expression.

Our computation is performed in two independent ways. We first neglect the light quark masses (i.e., for $q_u=u,c$ and $q_d=s,d,b$) and expand in the external momenta before integration~\cite{Smirnov:1994tg}
\begin{equation}
\dfrac{1}{(k+p)^2-M^2} = \dfrac{1}{k^2-M^2}\left[1-\dfrac{p^2+ 2 \left( k\cdot p\right)}{k^2-M^2}+ \dfrac{4 \left( k\cdot p\right)^2}{(k^2-M^2)^2}\right]+\mathcal{O}\left(\dfrac{p}{M}\right)^4\,,
\end{equation}
\noindent where $k$ is the loop momentum, $p$ is a generic external momentum and $M$ stands for the mass of the particle running in the loop. With this method we obtain analytic expressions for the loop functions, systematically accounting for the corrections of order $\mathcal{O}(m_Z/m_{\Delta})^n$ (with $n>0$), but avoiding the difficult computation of Passarino-Veltman functions with nonzero external momenta. We then compare these expressions with the ones computed numerically by using the Mathematica packages LoopTools~\cite{Hahn:1998yk} and Package-X~\cite{Patel:2015tea}. We find an agreement better than per-mil level between the results obtained numerically and analytically, for leptoquark masses heavier than $\approx 900$~GeV, as currently allowed by LHC searches~\cite{Angelescu:2018tyl}.

We present now our analytic results in terms of the Yukawas $l_{ij}$ and $r_{ij}$, and the quantum numbers of SM fermions. As explained above, we separate the top quark contribution from the light quarks ($q_u=u,c$ and $q_d=s,d,b$), since the relevant scales are different in each case. The final result for $F=0$ leptoquarks to $\mathcal{O}(m_Z^2/m_{\Delta}^2)$ reads 

\begin{align}
\label{eq:g-F0-final}
\begin{split}
\Big{[}\delta g_{\ell_{L(R)}}^{ij} \Big{]}_{F=0} =  &N_C\,\dfrac{w_{tj}w_{ti}^\ast}{16 \pi^2} \Bigg{[} \left(g_{u_{L(R)}}-g_{u_{R(L)}}\right)\dfrac{x_t (x_t-1- \log x_t)}{(x_t-1)^2}+ {\dfrac{x_Z}{12}} F_{F=0}^{L(R)}(x_t)\Bigg{]}\\
&+x_Z\, N_C  \sum_{k=u,c}\dfrac{w_{kj}w_{ki}^\ast}{48 \pi^2}  \Bigg{[} -g_{u_{R(L)}} \left(\log x_Z-i\pi -\dfrac{1}{6}\right)+\dfrac{g_{\ell_{L(R)}}}{6}\Bigg{]}\\
&+x_Z\, N_C  \sum_{k=d,s,b}\dfrac{w_{kj}w_{ki}^\ast}{48 \pi^2}  \Bigg{[} -g_{d_{R(L)}} \left(\log x_Z-i\pi -\dfrac{1}{6}\right)+\dfrac{g_{\ell_{L(R)}}}{6}\Bigg{]}\,,
\end{split}
\end{align}

\noindent while for $|F|=2$ leptoquarks we obtain,

\begin{align}
\label{eq:g-F2-final}
\begin{split}
\Big{[}\delta g_{\ell_{L(R)}}^{ij} \Big{]}_{F=2} =    &N_C\,\dfrac{w_{tj}w_{ti}^\ast}{16 \pi^2} \Bigg{[} \left(g_{u_{L(R)}}-g_{u_{R(L)}}\right)\dfrac{x_t (x_t-1- \log x_t)}{(x_t-1)^2}+ {\dfrac{x_Z}{12}} F_{F=2}^{L(R)}(x_t)\Bigg{]}\\
&+x_Z\, N_C  \sum_{k=u,c}\dfrac{w_{kj}w_{ki}^\ast}{48 \pi^2}  \Bigg{[} g_{u_{L(R)}} \left(\log x_Z-i\pi -\dfrac{1}{6}\right)+\dfrac{g_{\ell_{L(R)}}}{6}\Bigg{]}\\
&+x_Z\, N_C  \sum_{k=d,s,b}\dfrac{w_{kj}w_{ki}^\ast}{48 \pi^2}  \Bigg{[} g_{d_{L(R)}} \left(\log x_Z-i\pi -\dfrac{1}{6}\right)+\dfrac{g_{\ell_{L(R)}}}{6}\Bigg{]}\,,
\end{split}
\end{align}

\noindent where $x_t=m_t^2/m_\Delta^2$, $x_Z=m_Z^2/m_\Delta^2$, $N_C=3$, with $g_L^f=I_3^f-Q_f \sin^2 \theta_W$ and $g_R^f = -Q_f \sin^2 \theta_W$  ($f=u,d,\ell$), as before. In the above expressions, $w_{ki}$ should be replaced by {$r_{ki}$} or {$l_{ki}$} for $\delta {g_L}^{ij}$ or $\delta {g_R}^{ij}$, respectively. These couplings are collected in Table~\ref{tab:LQ-couplings} for each leptoquark representation listed in Sec.~\ref{Sec:1}. One of the novelties of our study is the inclusion of the terms $\mathcal{O}(x_Z \log x_t)$, which have never been considered before and which can induce non-negligible corrections for LQ masses $m_\Delta \lesssim 1.5$~TeV.~\footnote{Note, in particular, that much lower masses are allowed by current LHC searches for leptoquarks, which exclude masses of order $\approx 900~\mathrm{GeV}$ for LQs with mostly couplings to the third generation, see e.g.~Ref.~\cite{Angelescu:2018tyl} for a recent review.} These corrections are collected in the functions $F_{0}^{L(R)}$ and $F_{2}^{L(R)}$, which are given by

\begin{align}
\begin{split}
F_{F=0}^{L(R)}(x_t) =  & g_{u_{R(L)}} \dfrac{(x_t-1)(5 x_t^2-7 x_t+8)-2(x_t^3+2)\log x_t}{(x_t-1)^4} \\&+ g_{u_{L(R)}} \dfrac{(x_t-1)(x_t^2-5 x_t-2)+ 6 x_t\log x_t}{(x_t-1)^4} \\&+ g_{\ell_{L(R)}} \dfrac{(x_t-1)(-11 x_t^2+ 7 x_t-2)+ 6 x_t^3\log x_t}{3(x_t-1)^4}  \,,
\end{split}
\end{align}

\noindent and 
\begin{align}
\begin{split}
F_{F=2}^{L(R)}(x_t) = &-g_{u_{L(R)}} \dfrac{(x_t-1)(5 x_t^2-7 x_t+8)-2(x_t^3+2)\log x_t}{(x_t-1)^4} \\&- g_{u_{R(L)}} \dfrac{(x_t-1)(x_t^2-5 x_t-2)+ 6 x_t\log x_t}{(x_t-1)^4}  \\&+ g_{\ell_{L(R)}} \dfrac{(x_t-1)(-11 x_t^2+ 7 x_t-2)+ 6 x_t^3\log x_t}{3(x_t-1)^4}  \,.
\end{split}
\end{align}
 
\noindent  The phenomenological relevance of these terms and the other finite contributions we computed for the first time will be illustrated in the following.

\

\begin{table}[htbp!]
\renewcommand{\arraystretch}{1.7}
\centering
\begin{tabular}{|c|c|c|cc|cccc|}
\hline 
Decay & $w_{ij}$ &$q$ &  $R_2$  & $\widetilde{R}_2$ & $S_1$ &  $S_3$  & $\widetilde{S_1}$ & $\bar{S}_1$\\ \hline\hline
\multirow{4}{*}{$Z\to \ell\ell$}  & \multirow{2}{*}{$r_{ij}$} & $q_u$ &  $-\Big{(}y_{R_2}^L\Big{)}_{ij}$     &  $0$ & $\Big{(}V^\ast y_{S_1}^L\Big{)}_{ij}$     	&  $-\Big{(}V^\ast y_{S_3}^L\Big{)}_{ij}$  & $0$ & $0$\\[0.3em] 
 & & $q_d$ &  $0$     & $-\Big{(}y_{{\widetilde{R}_2}}^L\Big{)}_{ij}$     & $0$	& $-\sqrt{2}\Big{(}y_{S_3}^L\Big{)}_{ij}$  & $0$& $0$\\[0.3em] \cdashline{2-9}
& \multirow{2}{*}{$l_{ij}$} & $q_u$ &  $\Big{(}V y_{R_2}^R\Big{)}_{ij}$      &  $0$ & $\Big{(}y_{S_1}^R\Big{)}_{ij}$ 	& $0$ & $0$ & $0$\\[0.3em]  
& &$q_d$ &    $\Big{(}y_{R_2}^R\Big{)}_{ij}$    &  $0$ & 	$0$ & $0$ & $\Big{(}y_{\widetilde{S}_1}^R\Big{)}_{ij}$ & $0$ \\[0.3em]
    \hline
    \hline 
\multirow{4}{*}{$Z\to \nu\nu$}  & \multirow{2}{*}{$r_{ij}$} & $q_u$ &   $\Big{(}y_{R_2}^L\Big{)}_{ij}$    & $0$    & 	$0$ & $\sqrt{2}\Big{(}V^\ast y_{S_3}^L\Big{)}_{ij}$& $0$ & $0$\\[0.3em] 
 & & $q_d$ &  $0$     &  $\Big{(}y_{{\widetilde{R}_2}}^L\Big{)}_{ij}$ & $-\Big{(}y_{S_1}^L\Big{)}_{ij}$ 	& $-\Big{(}y_{S_3}^L\Big{)}_{ij}$  &  $0$ & $0$\\[0.3em] \cdashline{2-9}
& \multirow{2}{*}{$l_{ij}$} & $q_u$ &  $0$    & $\Big{(}V y_{{\widetilde{R}_2}}^R\Big{)}_{ij}$& 	$0$ & $0$ & $0$ & $\Big{(}y_{\bar{S}_1}^R\Big{)}_{ij}$ \\[0.3em]  
& &$q_d$ &   $0$    & $\Big{(}y_{{\widetilde{R}_2}}^R\Big{)}_{ij}$ & $\Big{(}y_{S_1}^{\prime\,R}\Big{)}_{ij}$	& $0$ & $0$ & $0$\\ 
    \hline
\end{tabular}
\caption{ \sl \small Expressions for the coefficients $w_{ij}$ in Eq.~\eqref{eq:g-F0-final} and \eqref{eq:g-F2-final} obtained by the matching of Eq.~\eqref{eq:yuk-F0} and \eqref{eq:yuk-F2} onto the Yukawa Lagrangians listed in Sec.~\ref{Sec:1} for LQs with fermion number $F=0$ and $|F|=2$, respectively. }
\label{tab:LQ-couplings} 
\end{table}

\subsection{Relevance of the finite terms in $Z\to \ell\ell$}
\label{Sec:finite-pieces-Z}

We now discuss the relevance of the new contributions we computed, namely the $\mathcal{O}(x_Z \log x_t)$ terms and the finite terms in the matching. To this end, we compare our results to the formulas given in Ref.~\cite{Feruglio:2016gvd}, obtained in a EFT context by employing a RGE approach to a leading-logarithmic approximation (LLA). The latter approach only accounts for the terms $x_t \log x_t$ and $x_Z \log x_Z$ from the general expressions, where one assumes that $v_{\mathrm{EW}} \approx m_Z \approx m_t$ in the logarithms. 

\begin{figure}[tp!]
\centering
\includegraphics[width=0.49\linewidth]{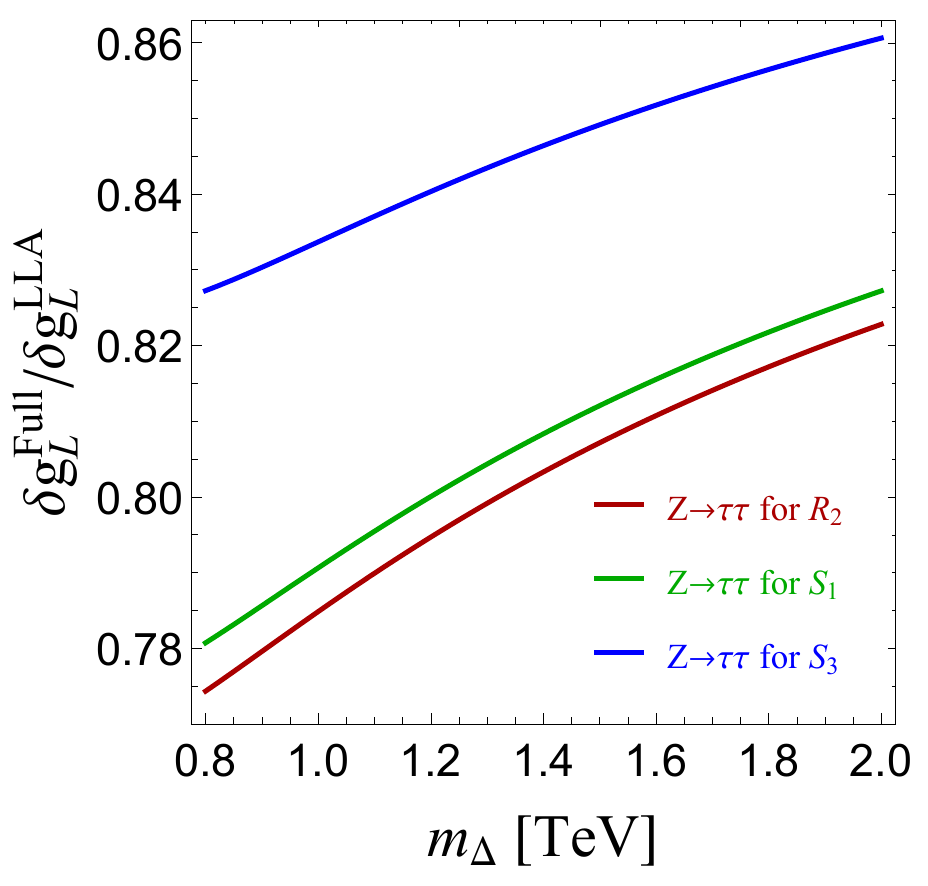} \includegraphics[width=0.49\linewidth]{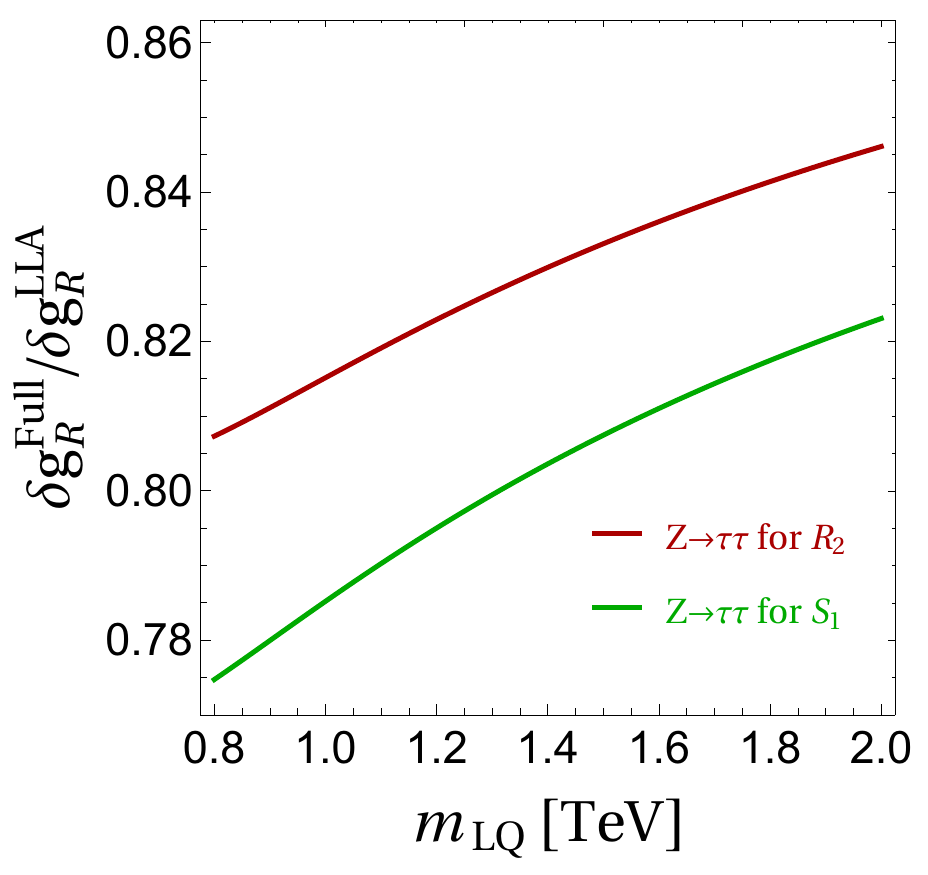} 
\vspace{8pt}

\includegraphics[width=0.49\linewidth]{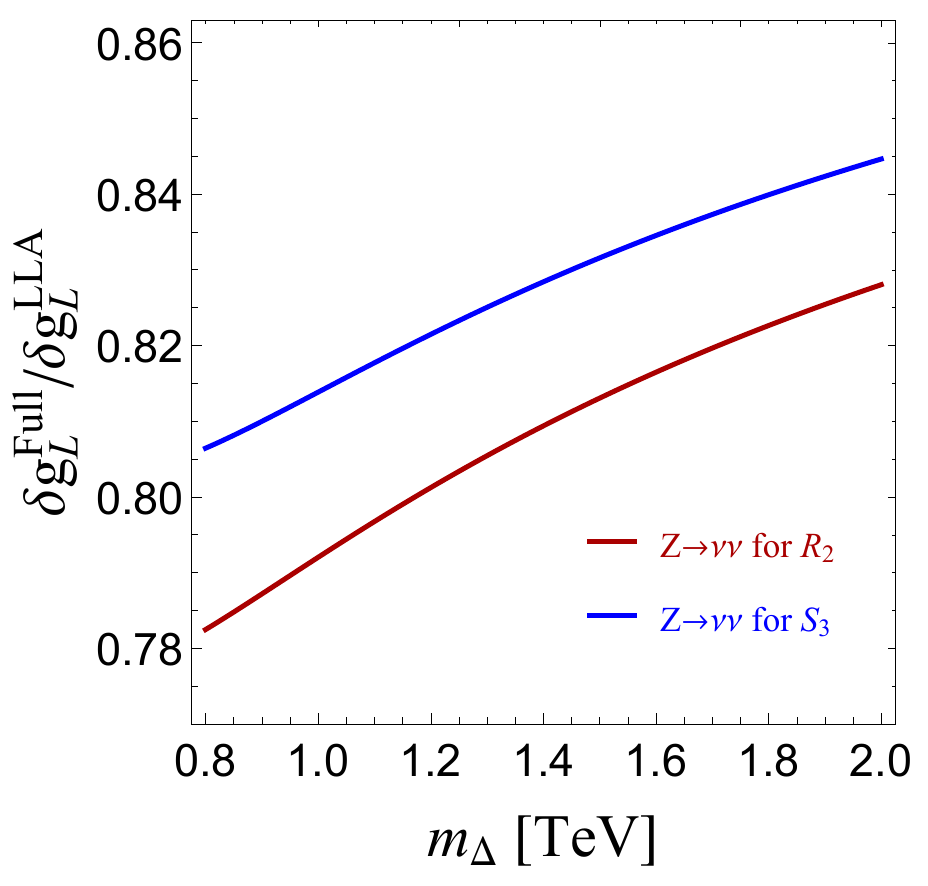} \includegraphics[width=0.49\linewidth]{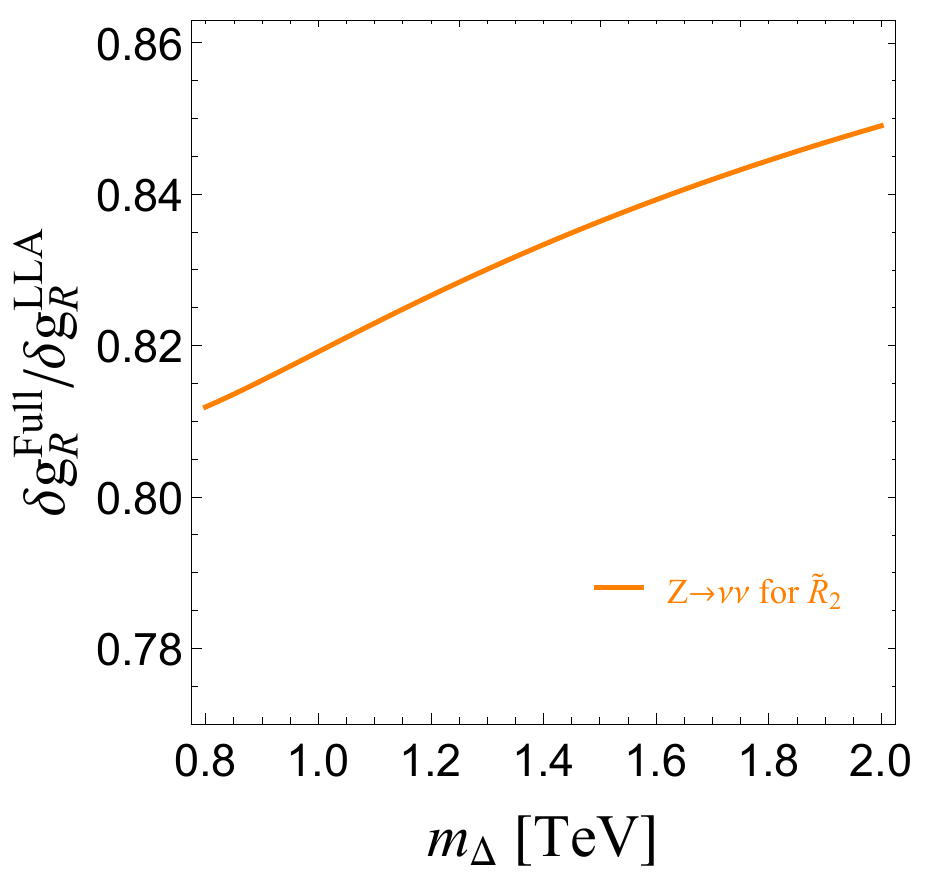} 
\caption{\small \sl Comparison of the full expressions given in Eq.~\eqref{eq:g-F0-final} and \eqref{eq:g-F2-final} for $Z\to \ell^+\ell^-$ and $Z\to \nu \bar{\nu}$ with the ones obtained by employing a RGE approach with a leading logarithmic approximation (LLA)~\cite{Feruglio:2016gvd}. We consider all LQ representations listed in Sec.~\ref{Sec:1} and we assume, for illustration, that the LQs only have couplings to third generation fermions in Eq.~\eqref{eq:yuk-R2}--\eqref{eq:yuk-S1bar}. The contributions from the new terms we have computed can be as large as $\mathcal{O}(20\%)$, for masses allowed by the direct searches at the LHC, being therefore non-negligible in phenomenological analyses.}
\label{fig:lla-comparison}
\end{figure}

In Fig.~\ref{fig:lla-comparison}, we show the ratio between the full and simplified formulas for $Z\to \ell^-\ell^+$ and for $Z\to \nu\bar{\nu}$ as a function of the LQ mass, for the different $SU(2)_L \times U(1)_Y$ representations. For illustration, we have only considered Yukawa couplings to third generation fermions in Eq.~\eqref{eq:yuk-R2}--\eqref{eq:yuk-S1bar}. We find that the new corrections we have computed can be as large as $\mathcal{O}(20\%)$ for values of LQ mass below $1.5$~TeV, as allowed by the present limits from the direct searches at the LHC~\cite{Angelescu:2018tyl}. Furthermore, we see that these relative corrections decrease with the LQ mass, becoming less relevant for larger masses, in which case the LLA is satisfied to a good extent. The conclusion of this exercise is that, given the present limits from LHC, one should consider the full formulas to reliably assess the viability of any scenario with low-energy scalar LQ. We will illustrate this feature in Sec.~\ref{Sec:3} with a concrete model for the $B$-anomalies, which presents a tension with current data if the formulas from Ref.~\cite{Feruglio:2016gvd} are used, but which turns out to be perfectly consistent when the full formulas are considered.

Before closing this Section we need to compare our results with previous computations in the literature. We agree with the results from Ref.~\cite{Bauer:2015knc}, where the light fermion and top-quark contributions have been computed for the $S_1$ model. To that result we included terms of $\mathcal{O}(x_Z \log x_t)$, which amount to a $\mathcal{O}(10\%)$ relative effect. We also agree with the results presented in Ref.~\cite{Becirevic:2017jtw}, where the $R_2$ contribution was computed for both light and heavy fermions. Again, our result goes a step beyond in that we include also  the $\mathcal{O}(x_Z \log x_t)$ terms. Similarly, if we neglect the $\mathcal{O}(x_Z \log x_t)$ terms, 
we agree with the results of Ref.~\cite{ColuccioLeskow:2016dox} where the top-quark contribution was computed for $S_1$ and $R_2$ LQs. Note, however, that we disagree with their sign of $\delta g_{L(R)}$ for the $R_2$ LQ. Before moving on, we stress that the expressions given above can be easily adapted to other scenarios of new physics containing scalar particles coupled to fermions, such as the two-Higgs doublet models~\cite{Branco:2011iw}.

\section{Leptoquark contributions to $W\to \ell\bar{\nu}$}
\label{Sec:3}

\subsection{Effective field theory description}
We now turn to the $W$ couplings to leptons. Similar to the above discussion, the $W$ interactions can be generically written as
\begin{align}
\label{eq:leff-w}
\delta \mathcal{L}_{\mathrm{eff}}^W = \dfrac{g}{\sqrt{2}} \sum_{i,j} \bar{\ell}_{i} \gamma^\mu
\Big{[}\left(\delta^{ij}+h_{\ell_L}^{ij}\right) \, P_L+  \delta h_{\ell_R}^{ij} \,P_R \Big{]} \nu_j\, W_\mu^- + \mathrm{h.c.\,,}\end{align}

\noindent where $h_{\ell_{L,R}}^{ij}$ describes the loop-level contributions illustrated in Fig.~\ref{fig:diagrams-Z}. In this expression, we also consider the possibility of light right-handed neutrinos, which we assume to be Dirac particles for simplicity.~\footnote{For $W$ decays, the expression for Majorana neutrinos is a trivial extension of the results presented above (cf. e.g. Ref.~\cite{Abada:2016plb} for further discussion).} The corresponding branching ratio can then be written as
\begin{equation}
\mathcal{B}(W\to \ell_i \nu_j) = \dfrac{m_W^3}{12 \pi v^2 \Gamma_W}\Big{(} |\delta_{ij}+\delta h_{\ell_L}^{ij}|^2 + |\delta h_{\ell_R}^{ij}|^2 \Big{)}\left(1- \dfrac{m_i^2}{2 m_W^2}-\dfrac{m_i^4}{2 m_W^4}\right)\,,
\end{equation}

\noindent where $m_i \equiv  m_{\ell_i}$ and neutrinos masses have been neglected. This expression should be compared to the LEP measurements~\cite{Tanabashi:2018oca}
\begin{align}
\mathcal{B}(W\to \tau \bar{\nu})^{\mathrm{exp}} = 11.38(21)\times 10^{-2}\,,\\[0.3em]
\mathcal{B}(W\to \mu \bar{\nu})^{\mathrm{exp}} = 10.63(15)\times 10^{-2}\,,\\[0.3em]
\mathcal{B}(W\to e \bar{\nu})^{\mathrm{exp}} = 10.71(16)\times 10^{-2}\,.
\end{align}

\noindent In particular, the ratio $R^{\tau/\mu}_W=\mathcal{B}(W\to \tau \nu)^{\mathrm{exp}}/\mathcal{B}(W\to \mu \nu)^{\mathrm{exp}}=1.07(3)$ is about $2.4 \sigma$ above the SM prediction, $R^{\tau/\mu,\,\mathrm{SM}}_W \approx 0.999$. It is very challenging to explain such a large deviation in a new physics model, since these contributions would be correlated, via $SU(2)_L \times U(1)_Y$ gauge invariance, with the tightly measured $Z$ couplings to $\tau$-leptons~\cite{Filipuzzi:2012mg}. Alternatively, the $W\tau \nu$ coupling can also be inferred from the $\tau$-lepton decays. Current PDG average~\cite{Tanabashi:2018oca}
\begin{equation}
\mathcal{B}(\tau \to \mu \nu\bar{\nu})^{\mathrm{exp}}=17.33(5)\times 10^{-2}\,,
\end{equation}

\noindent in a good agreement with the SM prediction, $\mathcal{B}(\tau\to\mu\nu\bar{\nu})=17.29(3)\times 10^{-2}$~\cite{Becirevic:2016zri}. Leptoquarks would also contribute to $\tau$-decays via box-type diagrams. Since these contributions are proportional to $y_{\mathrm{LQ}}^4/m_{\Delta}^2 = m_{\Delta}^2 \times \left(y_{\mathrm{LQ}}/m_{\Delta}\right)^4$, where $y_{\mathrm{LQ}}$ denotes a generic LQ coupling, we know these are subdominant contributions for low values of $m_{\Delta}$ and fixed values of $y_{\mathrm{LQ}}/m_{\Delta}$, as in the case of the $B$-anomalies.  For completeness we provide the expression for $\mathcal{B}(\tau\to\mu\nu\bar{\nu})$ in Appendix~\ref{sec:taumununu} which will be further discussed on one of our future publications.

\begin{figure}[ht!]
\centering
\includegraphics[width=0.55\linewidth]{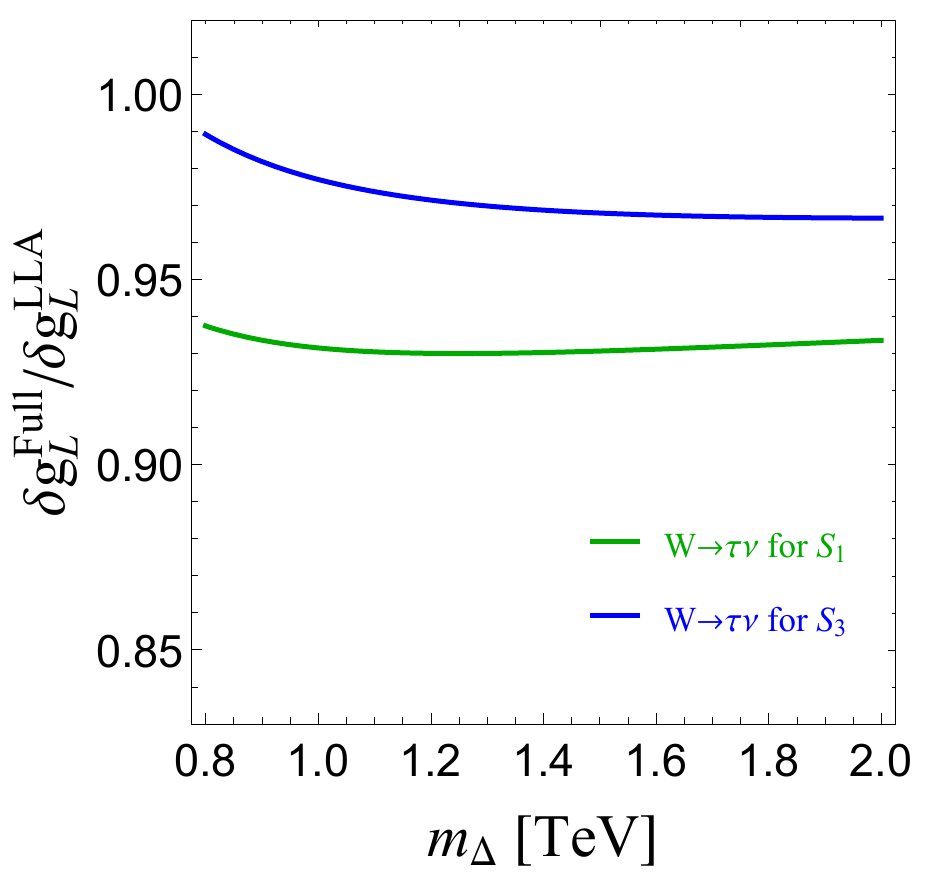}
\caption{\small \sl Comparison of the full expressions given in Eq.~\eqref{eq:WS1} and \eqref{eq:WS3} for $W\to \ell\bar{\nu}$ with the ones obtained by employing a RGE approach with a leading logarithmic approximation (LLA)~\cite{Feruglio:2016gvd}. For illustration, we assume once again that the LQs only have couplings to the third generation fermions in Eq.~\eqref{eq:yuk-R2}--\eqref{eq:yuk-S1bar}. We do not display the results for the $R_2$ and $\widetilde{R}_2$ LQs, since the RGE approach cannot encapsulate the contributions from these states, which have no logarithmic dependence. Furthermore, note that the LQs $\bar{S}_1$ and $\widetilde{S}_1$ do not contribute to these decays to one-loop order.
}
\label{fig:lla-comparison-W}
\end{figure}

\subsection{One-loop matching}

We now give the expressions for $h_{\ell_L}^{ij}$ and $h_{\ell_R}^{ij}$ for each LQ model listed in Sec.~\ref{Sec:1}. From Eq.~\eqref{eq:yuk-S1t} and \eqref{eq:yuk-S1bar}, we see that the models $\widetilde{S_1}=\left( \overline{\mathbf{3}}, \mathbf{1}\right)_{4/3}$ and $\overline{S}_1=\left( \overline{\mathbf{3}}, \mathbf{1}\right)_{-2/3}$ do not contribute to $W\to \ell \bar{\nu}$, since these are singlets of $SU(2)_L$ which do not have couplings to both up- and down-type quarks, neither to the $W$. For the scenarios with weak doublet leptoquarks, we obtain 
\begin{align}
\label{eq:WR2}
\Big{[}\delta h_{\ell_L}^{ij}\Big{]}_{R_2} &= N_{\rm c}
 \frac{x_W}{288 \pi^2} 
\Big{[}\left( y^{L\dagger}_{R_2}\right)_{it} \left(y^{L}_{R_2}\right)_{tj}G_{R_2}(x_t)
+\sum_{k=u,c}\left( y^{L\dagger}_{R_2}\right)_{ik} \left(y^{L}_{R_2}\right)_{kj}\Big{]}\,, \\[0.3em]
\label{eq:WR2tilde}
\Big{[}\delta h_{\ell_L}^{ij}\Big{]}_{\widetilde{R_2}} &= N_{\rm c} 
   \frac{x_W}{288 \pi^2} \left( y^{L\dagger}_{\widetilde{R}_2} y^{L}_{\widetilde{R}_2}\right)_{ij} \,,
\end{align}
\noindent where $x_W=m_W^2/m_{\Delta}^2$ and $x_t = m_t^2/m_{\Delta}^2$, as before, and the function $G_{R_2}$ is defined by
\begin{align}
G_{R_2}(x_t) = &\frac{-11 x_t^3+6 x_t^3 \log x_t+18 x_t^2-9 x_t+2}{2 (x_t-1)^4}\,,
\end{align}

\noindent Note, in particular, that these contributions cannot be accounted for by the EFT computation with leading-logarithmic approximation~\cite{Feruglio:2016gvd}. For the two remaining scenarios, we find

\begin{align}
\begin{split}
\Big{[}\delta h_{\ell_L}^{ij}\Big{]}_{S_1} = &N_C\Big{(}V^\ast y_{S_1}^L\Big{)}_{ti}^\ast \Big{(}V^\ast y_{S_1}^L\Big{)}_{tj}\Bigg{[}-\dfrac{x_t (x_t-1+(x_t-2)\log x_t)}{64 \pi^2(x_t-1)^2}
+\dfrac{x_W}{288 \pi^2} G_{S_1}(x_t)  \Bigg{]} \\[0.3em]
&+ N_C \dfrac{x_W}{144 \pi^2}\sum_{k=u,c} \Big{(}V^\ast y_{S_1}^L\Big{)}_{ki}^\ast \Big{(}V^\ast y_{S_1}^L\Big{)}_{kj} \left(-1-3 \log x_W +3 \pi i\right)  \,,
\end{split}
\label{eq:WS1}
\end{align}
and
\begin{align}
\begin{split}
\Big{[}\delta h_{\ell_L}^{ij}\Big{]}_{S_3} = &N_C\Big{(}V^\ast y_{S_3}^L\Big{)}_{ti}^\ast \Big{(}V^\ast y_{S_3}^L\Big{)}_{tj} \Bigg{[}\dfrac{x_t (x_t-1+(x_t-2)\log x_t)}{64 \pi^2(x_t-1)^2}+\dfrac{x_W}{288 \pi^2} G_{S_3} (x_t)\Bigg{]}  \\[0.3em]
&+ N_C \dfrac{x_W}{144 \pi^2}\sum_{k=u,c} \Big{(}V^\ast y_{S_3}^L\Big{)}_{ki}^\ast \Big{(}V^\ast y_{S_3}^L\Big{)}_{kj} \left(1-3 \log x_W +3 \pi i\right)  \,,
\end{split}
\label{eq:WS3}
\end{align}
\noindent where we separate the top-quark contributions from the other light quarks. The functions $G_{S_1}$ and $G_{S_3}$  are given by
\begin{align}
 G_{S_1}(x_t) =&\frac{6( x_t-1-\log x_t)}{(x_t-1)^2}\,,\\
G_{S_3} (x_t) = &\frac{6 \left[x_t \left(x_t^2+x_t-2\right)+1\right] \log
   x_t+x_t-1 [x_t (x_t (2
   x_t-23)+25)-10]}{(x_t-1)^4}\,.
\end{align}
Finally, note that none of the scalar LQ particles contribute at one-loop order to $h_{\ell_R}^{ij}$. 

\subsection{Relevance of the finite terms in $W\to \ell\nu$}
\label{Sec:finite-pieces-W}

We should now comment on the phenomenological implications of the results presented above. Similarly to the discussion of leptonic $Z$ couplings in Sec.~\ref{Sec:2}, we compared our full formulas to the ones obtained within a leading logarithmic approximation \cite{Feruglio:2016gvd}. These results are illustrated in Fig.~\ref{fig:lla-comparison-W} for the models $S_1$ and $S_3$, where we considered couplings only to the third generation fermions. For both scenarios, we find a negative correction coming from finite terms of order $\mathcal{O}(5\%)$, with a very mild dependence on the LQ mass $m_\Delta$. For the other scenarios, namely $R_2$ and $\widetilde{R}_2$, we cannot perform such a comparison since the leading logarithmic approximation of Eq.~\eqref{eq:WR2} and \eqref{eq:WR2tilde} would give a vanishing contribution. In this case, the finite terms are essential to consider.

\section{Illustration: $S_1$ $\&$ $S_3$ explanation of $R_{K^{(\ast)}}$ and $R_{D^{(\ast)}}$}
\label{Sec:4}

In this Section we illustrate our results in a specific scalar LQ model proposed to simultaneously explain the $b\to s$ and $b\to c$ anomalies~\cite{Crivellin:2017zlb,Buttazzo:2017ixm,Marzocca:2018wcf}. This model contains the LQs $S_1 = \left( \overline{\mathbf{3}}, \mathbf{1}\right)_{1/3}$ and $S_3 = \left( \overline{\mathbf{3}}, \mathbf{3}\right)_{1/3}$, with couplings only to left-handed fermions, namely
\begin{align}
\mathcal{L}_{\mathrm{Yuk}} &= \left( y_{S_1}^L\right)_{ij}\, \bar{Q}^C_i i\tau_2 S_{1} L_{j}+\left( y_{S_3}^L\right)_{ij}\,  \bar{Q}^C_i i\tau_2 (\vec{\tau} \cdot \vec{S}_{3})  L_{j} +\textrm{h.c.}\,.
\end{align}
 We adopt the same Yukawa pattern of Ref.~\cite{Buttazzo:2017ixm,Marzocca:2018wcf}, namely

\begin{equation}
\label{eq:texture-s1-s3-lh}
y^{L}_{S_1} =g_{S_1}\times
\begin{pmatrix}
  0 & 0 & 0\\
  0 & \beta_{s \mu}^{S_1} & \beta_{s \tau}^{S_1}\\
  0 & \beta_{b \mu}^{S_1} & \beta_{b \tau}^{S_1}
\end{pmatrix}\,,~\quad\qquad
y^{L}_{S_3} = g_{S_3} \times
\begin{pmatrix}
  0 & 0 & 0\\
  0 & \beta_{s \mu}^{S_3} & \beta_{s \tau}^{S_3}\\
  0 & \beta_{b \mu}^{S_3} & \beta_{b \tau}^{S_3}
\end{pmatrix}\,,
\end{equation}

\noindent where $g_{S_{1(3)}}$ describe the overall strength of LQ Yukawa interactions, while $\beta_{ij}^{S_{1(3)}}$ contain the flavor structure. Couplings to the first generation are set to zero to avoid stringent bounds from kaon physics observables and atomic parity violation. Following Ref.~\cite{Buttazzo:2017ixm,Marzocca:2018wcf}, we further assume that $\beta^{S_1}_{b\tau}=\beta^{S_3}_{b\tau}=1$, and that $\beta_{q\mu}\equiv\beta^{S_1}_{q\mu}=\beta^{S_3}_{q\mu}$, with $q=s,b$. The $s\tau$ couplings are considered to be in general different, as needed to explain current deviations. We are then left with six couplings to be fixed by the data, namely $g_{S_1}$, $g_{S_3}$, $\beta_{s\mu}$, $\beta_{b\mu}$ and $\beta_{s\tau}^{S_{1(3)}}$, as well as two masses $m_{S_1}$ and $m_{S_3}$.

Several low-energy observables are sensitive to the couplings defined above. To illustrate the impact of the expressions we computed for the first time in this paper, we consider the same experimental constraints of Ref.~\cite{Buttazzo:2017ixm}: (i) the LFU ratios $R_{K^{(\ast)}}$ and $R_{D^{(\ast)}}$, (ii) LFU tests in $R_D^{(\mu/e)}=\mathcal{B}(B\to D \mu \bar{\nu})/\mathcal{B}(B\to D e \bar{\nu})$, (iii) limits on the branching fractions $\mathcal{B}(B\to K^{(\ast)}\nu\bar{\nu})$, and (iv) the decays $Z\to \tau\tau$ and $Z\to \nu \bar{\nu}$. Concerning the latter observables, we perform a fit by using the leading-log approximation (LLA) of Ref.~\cite{Feruglio:2016gvd}, which is also considered in Ref.~\cite{Buttazzo:2017ixm,Marzocca:2018wcf}, and by considering our complete formulas, cf.~Sec.~\ref{Sec:2}. We consider the same range of parameters as in Ref.~\cite{Buttazzo:2017ixm}, namely $\beta_{s\mu},\beta_{s\tau}^{S_{1(3)}}\in (-5 \,V_{cb},5\, V_{cb})$ and $\beta_{b\mu}\in(-1,1)$.

\begin{figure}[ht!]
\centering
\includegraphics[width=0.7\linewidth]{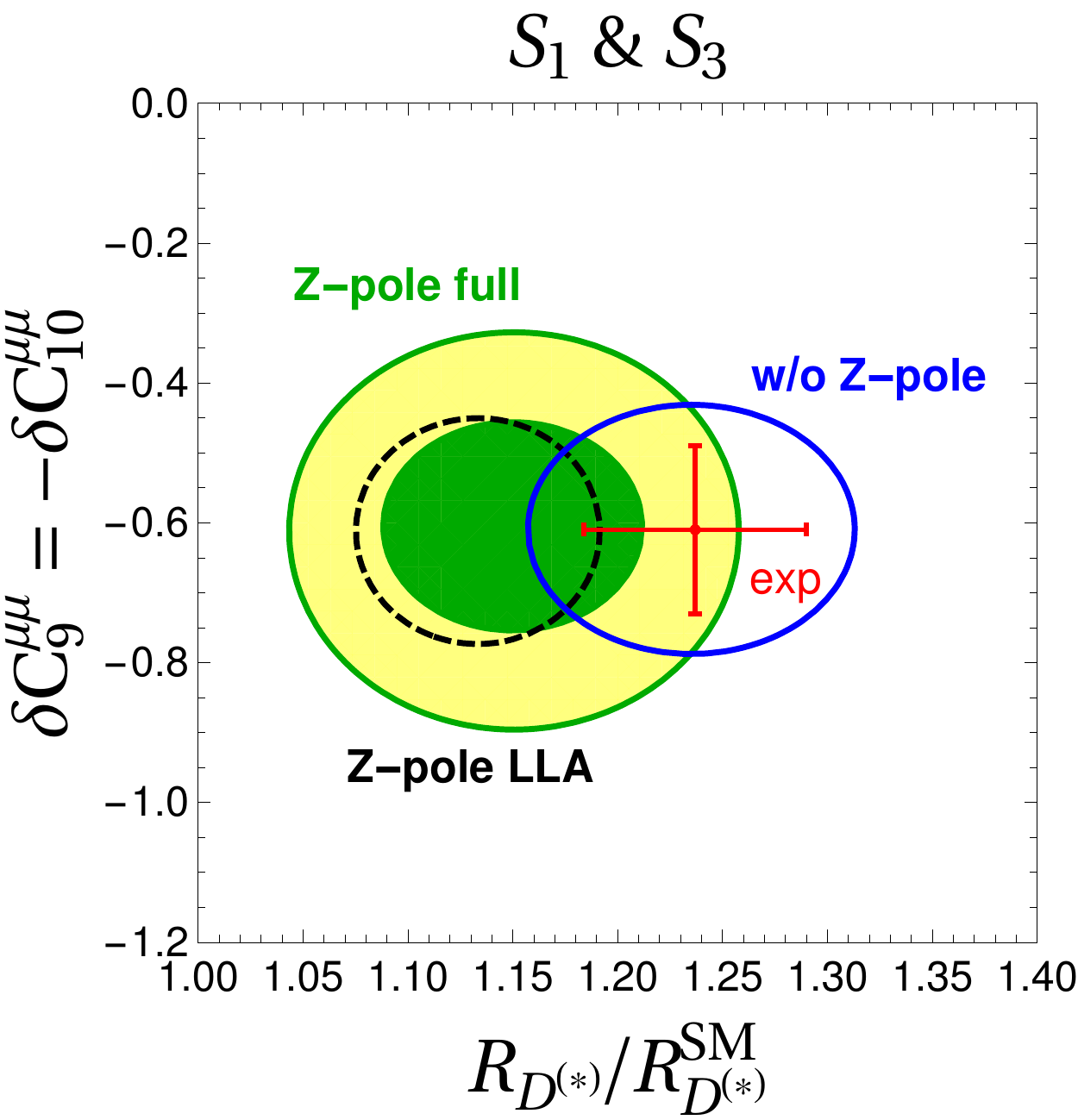}
\caption{\small \sl  Results of the low-energy fit considering the full expressions computed in this paper are depicted in the plane $R_{D^{(\ast)}}/R_{D^{(\ast)}}^{\mathrm{SM}}$ vs.~$\delta C_9^{\mu\mu}=-C_{10}^{\mu\mu}$ by the green (yellow) regions to $1\sigma$ ($2\sigma$) accuracy. Leptoquark masses are fixed to $m_{S_1}=m_{S_3}=2~\mathrm{TeV}$ and Yukawa couplings are scanned over the ranges described in the text. In the same plot we show the $1\sigma$ results of the fit by neglecting $Z$-pole constraints (blue line) and by including $Z$-poles observables with a leading-log approximation (black dashed line)~\cite{Feruglio:2016gvd}. As discussed in the text, the inclusion of finite terms reduces the tension between $R_{D^{(\ast)}}^{\mathrm{exp}}$ and $Z$-pole data. 
}
\label{fig:S1-S3-plot}
\end{figure}

Our results are depicted in Fig.~\ref{fig:S1-S3-plot} in the plane $R_{D^{(\ast)}}/R_{D^{(\ast)}}^{\mathrm{SM}}$ vs.~$\delta C_9^{\mu\mu}=-C_{10}^{\mu\mu}$  for LQ masses $m_{S_1}=m_{S_3}=2~\mathrm{TeV}$. In the analysis considering the leading-log approximation, we obtain a value $\chi_{\mathrm{min}}^2 \approx 8.0$, which shows a mild tension between the observed deviations $R_{D^{(\ast)}}$ and $Z$-pole data, as depicted by the $2\sigma$ contour (black dashed line). If, instead, one considers the formulas computed in this paper,  the tension is milder as shown by the green/yellow contours in Fig.~\ref{fig:S1-S3-plot}, for which we obtain $\chi_{\mathrm{min}}^2 \approx 6.5$. This example illustrates the importance of the finite one-loop terms in the computation of $Z\to \ell\ell$ and $Z\to \nu\nu$, which have a non-negligible effect for the models aiming at explaining the $B$-physics anomalies. Finally, it should be noted that similar conclusions have been reached in Refs.~\cite{Bordone:2017bld,Bordone:2018nbg} in which the authors considered an ultraviolet complete scenario which includes vector LQ states.

\section{\label{Sec:C} Summary and Conclusions}
In this paper we computed the radiative corrections of $Z$ and $W$ decays to two leptons/neutrinos induced by the scalar leptoquarks. 
We extend the leading logarithmic approximation $(\mathcal{O}(x_t \log x_t))$ by computing finite terms and, for the first time, we also include the terms $\mathcal{O}(x_{Z(W)}\log x_t)$. The results for $Z \to \ell \ell$ and $Z \to \nu \nu$ corrections are presented in a unified way, only separating the results for $F=0$ and $|F|=2$ leptoquarks. In this way one can easily compute the left and right-handed contributions using Table~\ref{tab:LQ-couplings}. These results can be easily extended to other models involving Yukawa couplings with a massive scalar. For the $W$ decays we present the results for each LQ separately. One remarkable feature of going further than the leading logarithmic approximation in the case of $W\to \ell \nu$ is that the $F=0$ LQ's bring in a non-vanishing contribution. In the appendix we also comment on $\ell_i\to \ell_j \nu_i \bar{\nu}_j$ and its relation to $W\to \ell \nu$.

The inclusion of finite terms and the terms containing $x_{Z(W)} \log x_t$ can change the $Z$ couplings to leptons and neutrinos by a 20\% while in the $W$ channel the modification amounts to a 5\% for LQ masses lower than 1.5 TeV. The 20\% difference in the $Z$ couplings is further illustrated on the example of a model of Ref.~\cite{Buttazzo:2017ixm} in which it is known that $Z\to \ell\ell$ creates a tension with $R_{D^{(\ast )}}/R_{D^{(\ast )}}^\mathrm{SM}$. We showed that the fit improves from $\chi^2_{\text{min}} \approx 8$ at LLA to $\chi^2_{\text{min}} \approx 6.5$  with our contributions.  This is also shown by Fig.~\ref{fig:S1-S3-plot} where the tension between 
the $Z-$pole observables and the $R_{D^{(*)}}$ anomalies is reduced if instead of LLA the results of our calculations are used.

\section*{Acknowledgments}
\label{sec:acknowledgments}
We thank Rupert Coy for pointing out a typo in the previous version. P.~A.~and F.~M.~are supported by MINECO grant
FPA2016-76005-C2-1-P and by Maria de Maetzu program grant
MDM-2014-0367 of ICCUB and 2017 SGR 929. This  project  has also  received  support  from  the  European  Union's  Horizon  2020  research  and  innovation
programme under the Marie Sklodowska-Curie grant agreement N.~690575 and 674896.

\clearpage


\appendix

\section{EFT description of $\ell_i \to \ell_j \nu_i \bar{\nu}_j $} \label{sec:taumununu}
\begin{figure}[t!]
\centering
\includegraphics[width=0.25\textwidth]{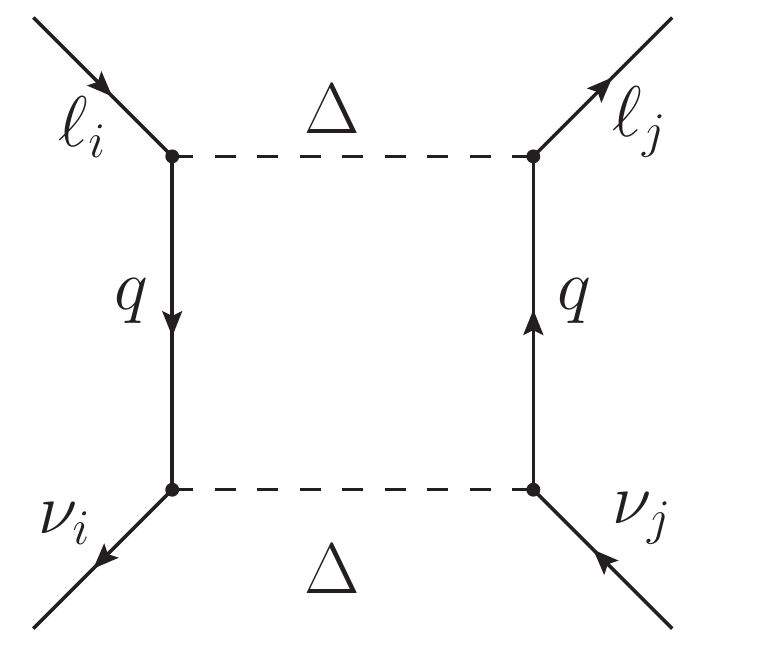}\quad\quad
\includegraphics[width=0.25\textwidth]{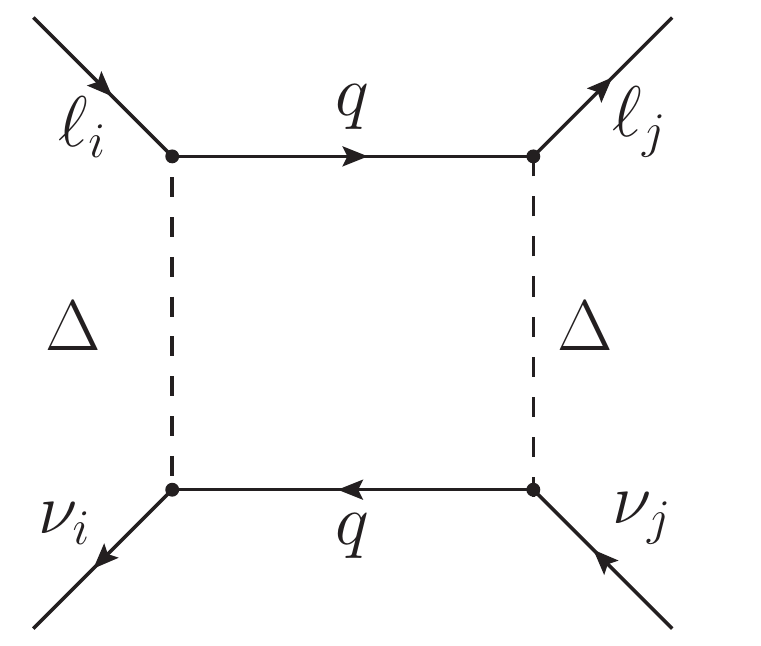}
\caption{\sl \small Box diagrams contributing to $\ell_i\to \ell_j \nu_i \bar{\nu}_j$  .}
\label{fig:boxtaumununu}
\end{figure}

In this Appendix we collect the complete expression for $\ell_i \to \ell_j \nu_i \bar{\nu}_j$, with $i,j\in\{e,\mu,\tau\}$, and $m_{\ell_i}>m_{\ell_j}$. The most general dimension six effective Lagrangian describing these decays without taking into account RH neutrinos can be written as
\begin{equation}
\begin{split}
\delta\mathcal{L}^\tau_\text{eff}=  -\frac{2}{v^2}\biggl[\left(1+\delta C_{LL}^{ij}\right) (\bar{\nu}_i \gamma^\mu P_{L} \ell_i)(\bar{\ell}_j \gamma_\mu P_{L} \nu_j)+ \delta C_{LR}^{ij} (\bar{\nu}_i P_{R} \ell_i)(\bar{\ell}_jP_{L} \nu_j)\biggr]+ \text{h.c.}\,, 
\end{split}
\end{equation}

\noindent where $\delta C_{LL}^{ij}$ and $\delta C_{LR}^{ij}$ are the Wilson coefficients. For simplicity, we have considered only the lepton flavor conserving couplings since the LFV ones would not interfere with the SM. The relevant decay rate then reads,
\begin{equation}
\label{eq:leff-taumununu}
\dfrac{\Gamma(\ell_i\to\ell_j \nu_i \bar{\nu}_j)}{\Gamma(\ell_i\to\ell_j \nu_i \bar{\nu}_j)^{\mathrm{SM}}} =  1+\left|1+\delta C_{LL}^{ij}\right|^2 + \dfrac{1}{4}\left|\delta C_{LR}^{ij}\right|^2 +\dfrac{2m_{\ell_i}}{m_{\ell_j}} \mathrm{Re}\biggl[(1+\delta C_{LL}^{ij})\delta C_{LR}^{ij}\biggr]\,,
\end{equation}

\noindent normalized with respect to the SM value, $\Gamma (\ell_i \to \ell_j \nu_i \bar{\nu}_j)^{\mathrm{SM}} = G_F^2 m_{\ell_i}^5/(192 \pi^3)$, after neglecting the terms $\mathcal{O}(m_{\ell_j}^2/m_{\ell_i}^2)$.  LQs contributes to the effective Wilson coefficients in Eq.~\eqref{eq:leff-taumununu} at the one-loop level via two types of diagrams: (i) $W$-penguins and (ii) box diagrams. The former ones can be expressed in terms of the $W\tau\nu$ effective vertices defined in Eq.~\eqref{eq:leff-w}. In the limit of small transferred momenta (i.e.~$m_{\ell_i}^2/m_W^2 \ll 1$) we find 
\begin{equation}
\label{eq:CLL-Wpenguin}
\Big{[}\delta C_{LL}^{ij}\Big{]}^{\mathrm{W-penguin}} = \delta h_{\ell_L}^{ii}(x_W=0)+\delta h_{\ell_L}^{jj}(x_W=0)+\dots\,,
\end{equation}
where $\delta h_L^{ij}$ are the effective coefficients reported in Sec.~\ref{Sec:3}, in which $x_W$ should be set to zero. 
In practice we truncate the series and neglect all the terms represented by `{\sl dots}' in Eq.~\eqref{eq:CLL-Wpenguin}. 
The box diagram contributions are schematically illustrated in Fig.~\ref{fig:boxtaumununu}. For the LQ doublets, we find
\begin{align}
\Big{[}\delta C_{LL}^{ij}\Big{]}_{R_2}^{\mathrm{box}}&=\frac{N_C v^2 }{128\pi^2 m_{R_2}^2}
\left(y^{L\,\dagger}_{R_2} \cdot y^{L}_{R_2}\right)_{jj}  \left(y^{L\,\dagger}_{R_2} \cdot y^{L}_{R_2}\right)_{ii}\,,\\[0.5em]
\Big{[}\delta C_{LR}^{ij}\Big{]}_{R_2}^{\mathrm{box}}&=-\frac{N_C v^2}{64\pi^2 m_{R_2}^2}
 \left( y^{R\,\dagger}_{R_2} \cdot y^{R}_{R_2}\right)_{ji}  \left(y^{L\,\dagger}_{R_2} \cdot y^{L}_{R_2}\right)_{ij}\,,
\end{align}
and
\begin{align}
\Big{[}\delta C_{LL}^{ij}\Big{]}^{\mathrm{box}}_{\widetilde{R}_2}&=-\frac{N_C v^2 }{128\pi^2 m_{\widetilde{R}_2}^2}
\left(y^{L\,\dagger}_{\widetilde{R}_2} \cdot  y^{L}_{\widetilde{R}_2}\right)_{jj} \cdot \left(y^{L\,\dagger}_{\widetilde{R}_2} \cdot y^{L}_{\widetilde{R}_2}\right)_{ii}\,,
\end{align}
 
\noindent where $y_{\mathrm{LQ}}^{\dagger}\cdot y_{\mathrm{LQ}}$ denotes a matrix product. The coefficient $\delta C_{LR}$ is not generated by $\widetilde{R}_2$ because this LQ does not couple to the right-handed leptons in Eq.~\eqref{eq:yuk-R2tilde}. Also note that none of these box contributions can be captured by an EFT computation to leading logarithms. For the remaining LQ models, we find
\beq
\begin{split}
\Big{[}\delta  C_{LL}^{ij}\Big{]}_{S_1}^{\mathrm{box}} &=
+ \dfrac{N_C v^2}{128 m_{S_1}^2}  \left( y^{L\,\dagger}_{S_1} y^{L}_{S_1}\right)_{ji}  \left(y^{L\dagger}_{S_1}  y^{L}_{S_1}\right)_{ij}
\,,\\
\Big{[}\delta  C_{LR}^{ij}\Big{]}_{S_1}^{\mathrm{box}}&=-\frac{N_C v^2 }{64\pi^2 m_{S_1}^2} \left( y^{R\,\dagger}_{S_1} \cdot y^{R}_{S_1}\right)_{ji}  \left(y^{L\,\dagger}_{S_1} \cdot y^{L}_{S_1}\right)_{ij}\,.
\end{split}
\label{eq:CVLLtaumununuS1}
\eeq
and
\beq
\begin{split}
\Big{[}\delta  C_{LL}^{ij}\Big{]}_{S_3}^{\mathrm{box}} &=
 \dfrac{N_C v^2}{128 m_{S_3}^2}  \left( y^{L\,\dagger}_{S_3}\cdot y^{L}_{S_3}\right)_{ji}  \left(y^{L\,\dagger}_{S_3} \cdot  y^{L}_{S_3}\right)_{ij}+\dfrac{N_C v^2}{ 32 m_{S_3}^2}
\left(y^{L\,\dagger}_{S_3} \cdot y^{L}_{S_3}\right)_{jj}  \left(y^{L\,\dagger}_{S_3} \cdot y^{L}_{S_3}\right)_{ii}
\,.
\end{split}
\label{eq: CVLLtaumununuS3}
\eeq

\noindent  These contributions should be added to the ones, presented in Eq.~\eqref{eq:CLL-Wpenguin}.

\end{document}